\documentstyle[11pt,aaspp4,flushrt,epsf]{article}

\def\vsp{\vphantom{\Bigl(}}

\begin{document} 
%
\vphantom{p}
\vskip -55pt
\centerline{\hskip 2.7truein\hfill
  Astrophysical Journal, in press,}
\centerline{\hskip 2.7truein\hfill
  Vol. 541, September 20, 2000}
\vskip 15pt

\title{COMPTON SCATTERING IN ULTRA-STRONG MAGNETIC FIELDS:\\
NUMERICAL AND ANALYTICAL BEHAVIOR\\
IN THE RELATIVISTIC REGIME}

\author{Peter L. Gonthier}
\affil{Hope College, Department of Physics, 27 Graves Place, Holland, MI
49422-9000 \\
{\it gonthier@physics.hope.edu}}

\author{Alice K. Harding and Matthew G. Baring}
\affil{NASA - Goddard Space Flight Center, Laboratory for High Energy
Astrophysics \\
Greenbelt, MD 20771 \\
{\it harding@twinkie.gsfc.nasa.gov}, and {\it baring@twinkie.gsfc.nasa.gov}}

\author{Rachel M. Costello}
\affil{The College of Wooster, Department of Physics, 1189 Beall Ave,
Wooster, OH 44691 \\
{\it costelrm@acs.wooster.edu}}

\author{Cassandra L. Mercer}
\affil{The Colorado College, Department of Physics, 14 E. Cache La Poudre \\
Colorado Springs, CO 80903 \\
{\it c\_mercer@cc.colorado.edu}}

\begin{abstract} 
This paper explores the effects of strong magnetic fields on the
Compton scattering of relativistic electrons.  Recent studies of 
upscattering and energy loss by relativistic electrons that have
used the non-relativistic, magnetic Thomson cross section for resonant
scattering or the Klein-Nishina cross section for non-resonant
scattering do not account for the relativistic quantum effects of
strong fields ($ > 4 \times 10^{12}$ G).  We have derived a simplified
expression for the exact QED scattering cross section for the
broadly-applicable case where relativistic electrons move along the
magnetic field.  To facilitate applications to astrophysical models, we
have also developed compact approximate expressions for both the
differential and total polarization-dependent cross sections, with the
latter representing well the exact total QED cross section even at the
high fields believed to be present in environments near the stellar
surfaces of Soft Gamma-Ray Repeaters and Anomalous X-Ray Pulsars.  We
find that strong magnetic fields significantly lower the Compton
scattering cross section below and at the resonance, when the incident
photon energy exceeds $m_ec^2$ in the electron rest frame.  The cross
section is strongly dependent on the polarization of the final
scattered photon.  Below the cyclotron fundamental, mostly photons of
perpendicular polarization are produced in scatterings, a situation
that also arises above this resonance for sub-critical fields.
However, an interesting discovery is that for super-critical fields, a
preponderance of photons of parallel polarization results from
scatterings above the cyclotron fundamental.  This characteristic is
both a relativistic and magnetic effect not present in the Thomson or
Klein-Nishina limits.
\end{abstract}



\section{INTRODUCTION}

Recent observations are providing evidence for the existence of
isolated neutron stars having ultra-strong magnetic fields. Assuming
that the spin-down of isolated neutron stars is a result of
electromagnetic dipole radiation, the measured period and the period
derivative give the strength of the surface magnetic field as $B_o=6.4
\times 10^{19} (P \dot P )^{1/2}$ (Shapiro \& Teukolsky 1983 and Usov
\& Melrose 1995).  Typical radio pulsars have period and period
derivative distributions (from the Princeton Pulsar Catalogue: Taylor,
Manchester, \& Lyne 1993) suggesting magnetic field strengths between
$10^{11}$ and $10^{13} G$, with about two dozen pulsars having
spin-down fields greater than $10^{13} G$; the highest to date is $B_o=
1.1 \times 10^{14} G$, a product of the Parkes multi-beam survey
(Camilo et al. 1999). Although there have not been any radio pulsars
detected with a magnetic field much exceeding $B_o = 10^{14}
G$ (perhaps with the exception of the unconfirmed observation of SGR
1900+14; Shitov 1999; Shitov, Pugachev \& Kutuzov 2000), growing
evidence for a new class of isolated neutron stars with ultra-strong
magnetic fields ($B_o > 10^{14} G$) has come from the observations of
Soft Gamma-Ray Repeaters (SGRs) and Anomalous X-ray pulsars (AXPs).
The five known SGRs are transient sources that undergo repeated
outbursts of gamma rays and all have been associated with young ($t <
10^5 yr.$) supernova remnants.  Last year, Kouveliotou et al. (1998a)
detected a 7.47 s period in quiescent emission of SGR1806-20 in the
X-ray band.  Recently, Hurley et al.  (1999) detected a periodicity of
5.16 s for SGR1900+14 during a giant burst having an energy of $\sim
10^{45} erg$.  Kouveliotou et al.  (1999) also observed the same period
in the quiescent X-ray emission.  Assuming dipole radiation torques,
the measured period derivatives imply surface magnetic fields between
$10^{14} - 10^{15} G$, well above the quantum critical field, $B_{\rm cr}
= 4.4 \times 10^{13} G$.

Evidence for ultra-strong magnetic fields has also come from
observations of AXPs, a group of six or seven X-ray pulsars with
supersecond periods that exhibit anomalous characteristics in
comparison to the properties of accreting X-ray pulsars.  The lack of
optical counterparts and orbital Doppler shifts (Steinle et al. 1987,
Mereghetti et al. 1992 and Mereghetti \& Stella 1995) suggest that
these objects are isolated pulsars.  Their more-or-less steady
spin-down and young characteristic ages, $t < 10^5 yr.$ (Vasisht \&
Gotthelf 1997), support this assertion.  Several AXPs have been
associated with young supernova remnants, also suggesting neutron star
origin.  The AXPs are bright X-ray sources with luminosities, $L_X\sim
10^{35} erg/s$, far exceeding their spin-down luminosity.   This
energetics issue has motivated Thompson \& Duncan (1996) and Kulkarni
\& Thompson (1998) to suggest that, unlike rotation-powered pulsars,
the X-ray and particle emission in AXPs is powered by a decay of the
magnetic field in the stellar interior.

Various studies indicate that inverse Compton scattering (ICS) plays a
significant role in the magnetospheric physics of strongly magnetized
neutron stars.  Relativistic electrons accelerated above the polar cap
can Compton scatter off thermal radiation from the neutron star
surface, producing high energy gamma rays that can power pair
cascades.  Daugherty \& Harding (1989) found that in the presence of a
strong magnetic field, resonant scattering greatly increases electron
energy losses over those of non-resonant scattering, making Compton
scattering efficient even at lower temperatures.  Recent studies have
indicated the importance of resonant Compton scattering over curvature
radiation in pulsar polar cap acceleration models.  Luo (1996) and
Zhang et al.  (1997) observed that if the polar cap temperature and the
magnetic field are sufficiently high, the thickness of the accelerating
gap is limited more efficiently by pairs from Compton scattered photons
than by pairs from curvature radiation photons.  Harding \& Muslimov
(1998) considered ICS by the trapped, back flowing positrons and found
that the pairs from the ICS photons may cause surface acceleration gaps
to be unstable, forcing them to higher altitudes. These studies
indicate the evolving, critical role of Compton scattering in polar cap
models.

Compton scattering is also very important in SGR radiation models.  The
highly super-Eddington luminosities of the bursts ensures high
densities of both photons and particles, so that scattering will be a
critical factor.  Paczynski (1992) proposed that the lower scattering
cross section below the resonance in the strong magnetic field (for
photons in the perpendicular polarization mode) could allow
super-Eddington luminosities.  However, Miller et al. (1995) argued
that scattering into the parallel mode keeps the radiation pressure
high, and thus the effective Eddington luminosity lower, in hydrostatic
atmospheres.  In Thompson \& Duncan's (1995) model for the radiation
from SGR bursts, Compton scattering plays a critical role in
establishing equilibrium between pairs and photons and in the spectral
formation.  Compton scattering may also be important in the photon
splitting cascade model for SGR burst emission (Baring 1995, Harding,
Baring \& Gonthier 1996 and Harding, Baring \& Gonthier 1997).  The
issues raised and discussed by these papers all depend critically on
the polarization state and the angular distribution of the photons
involved in a scattering event.
 
The full QED expressions for the relativistic, magnetic cross section 
of Compton scattering were derived separately by Daugherty and Harding 
(1986, hereafter DH86), and Bussard et al. (1986) and discussed by
Meszaros (1992).  Because of the 
complexity of the expressions, they have been applied to the study 
of relativistic Compton scattering in high magnetic fields only for the 
case of a one-dimensional, thermal electron distribution.  These studies 
(Alexander \& Meszaros 1989, 1991, Harding \& Daugherty 1991, 
Araya \& Harding 1996, 1999) focussed on models of cyclotron line formation 
in accreting neutron star atmospheres and gamma-ray bursts.
The inverse Compton scattering models for pulsars and SGRs described in
the preceeding paragraphs involve non-thermal, highly relativistic
electrons.  Such models have not to date incorporated the QED scattering 
cross sections, because the larger length scales require treatment of
field inhomogeneity (unlike the cyclotron line scattering models which 
assume homogeneous fields).  Consequently, these inverse Compton scattering 
studies have had to approximate the scattering rates by a combination of 
the Klein-Nishina cross section for non-resonant Compton scattering and the non-relativistic (Thomson) limit (Canuto, Lodenquai, \& Ruderman 1971, 
Blandford \& Scharlemann 1976 and Herold 1979) for resonant Compton scattering. 
As a result, they do not include the quantum relativistic effects of a
strong magnetic field ($B > 0.1$, where here and throughout the paper $B$ 
is given in units of the critical field, $B_{\rm cr} = m_e^2c^3/e\hbar = 
4.414 \times 10^{13}$Gauss). 

It is the purpose of this paper to explicitly present the major
features of Compton scattering in strong sub-critical and
super-critical magnetic fields, providing a development of simplified
expressions for the magnetic scattering cross section, to facilitate
applications to environments near the surfaces of pulsars, SGRs or
AXPs.  We extend the work of DH86 by obtaining expressions suitable for
rapid computation of the polarized, differential and integrated cross
sections, applicable to Compton scattering of highly relativistic
electrons moving along the magnetic field.  In this case, broadly
applicable in astrophysical problems, photons propagate along the field
in the electron rest frame.  In this specialized axisymmetric case,
there is a degeneracy between polarization transitions $\perp
\rightarrow \parallel $ and $\parallel \rightarrow \parallel$, with a
similar identity of cross sections for the modes
$\perp\rightarrow\perp$ and $\parallel\rightarrow\perp$. Below the
cyclotron fundamental, mostly $\perp$-photons are produced in
scatterings, a situation that also pertains above this resonance for
sub-critical fields. However, an interesting discovery of this paper is
that for super-critical fields, the reverse situation arises above the
cyclotron fundamental, and a preponderance of photons of parallel
polarization results from scatterings.  We derive an analytic
approximation that describes well the integrated cross section for
Compton scattering in both sub-critical and super-critical magnetic
fields. The effects of the strong magnetic fields on the
angle-integrated cross sections and angular distributions of scattered
photons is also studied, noting in particular significant differences
for large scattering angles between the high energy magnetic forms and
the non-magnetic Klein-Nishina results.  Finally, we discuss the impact
of this field-dependent cross section on the simulation of acceleration
and cascade processes in pulsars and SGR sources.

\section{QED COMPTON SCATTERING CROSS SECTION}

The present study follows the development of DH86, applying their work
to a particular case for scattering of relativistic electrons.  The
expressions developed in this paper use the Johnson \& Lippmann (1949)
electron wavefunctions. Graziani (1993) and Graziani, Harding \& Sina
(1995) have indicated that the choice of these wavefunctions do not
diagonalize the magnetic moment operator parallel to the external field
resulting in some limitations and inaccuracies when spin-dependent
cross sections are used.  Therefore, in this work, we present results
that are spin-averaged.  We will further discuss this issue in section
4. The differential cross section in the rest frame of the electron is
given in equation (11) of DH86, with the denominator term later
corrected in Harding \& Daugherty (1991), by the expression
\begin{equation}
   {{d\sigma } \over {d\Omega '} } = { {3\sigma_{\rm T}} \over {16\pi } }
   {{\omega '\; e^{-(\omega '^2\sin ^2\theta ' +\omega^2\sin ^2\theta )/2B}}
      \over {\omega \left( {1+\omega -\omega '-(\omega \cos \theta
   -\omega '\cos \theta ')\cos \theta '} \right)\left( {2+\omega -\omega '}
   \right)}} \left| {\sum\limits_{n=0}^\infty  {\left[
   {F_n^{(1)}e^{i\Phi _1}+F_n^{(2)}e^{i\Phi _2}} \right]}} \right|^2.
\label{difforig}
\end{equation}
While photon-electron interactions may excite electrons above the
ground state in this case (to quantum numbers $\ell\geq 0$), the
initial electron is assumed to be in its ground state with $m=0$ Landau
state and its spin anti-parallel to the magnetic field.  This
assumption is valid as the synchro-cyclotron decay of excited states
occurs during an extremely short time-scale.  The incident and
scattered photon energies denoted by, $\omega$ and $\omega'$,
respectively are in units of the electron rest mass energy, $m_ec^2$.
The incident and scattered angles are denoted by $\theta$ and
$\theta'$, respectively, with respect to the z-axis determined by the
direction of the magnetic field. The common phase factors are given by
\begin{equation}
   \Phi _1=-\Phi _2={{\omega \omega '\sin \theta \sin \theta '} \over
   {2B}}\sin (\phi -\phi ').
\end{equation}
The sum in equation~(\ref{difforig}) is over the intermediate Landau
states, labelled by quantum number $n$.  The $F$ terms and phase
factors are associated with the two different Feynman diagrams shown in
Figure 1, and are listed explicitly in the Appendix.  The $F$ terms of
the second Feynman diagram can be obtained from those of the first
diagram through the crossing-symmetry replacements
\begin{equation}
   \omega \Leftrightarrow -\omega ',\ \ \ k\Leftrightarrow -k',\ \ \beta
   \Leftrightarrow \beta ',\ \hbox{and}\
   \varepsilon \Leftrightarrow \varepsilon'^*  \quad .
 \label{eq:cross-symm}
\end{equation}
For each Feynman diagram, each $F$ expression has a spin no-flip and a
spin flip $F$ term associated with it, due to the spin degeneracy of
the final states.  Notation used in defining the $F$ terms and
appearing elsewhere in this paper includes $\varepsilon$, which
represents the photon polarization components defining two orthogonal
linear polarization vectors as given in Daugherty \& Bussard (1980):
\begin{eqnarray}
   \varepsilon _{}^{\parallel} & = & -\cos \theta \cos \phi \hat x-\cos
\theta \sin \phi \hat y+\sin \theta \hat z, \nonumber\\
   \varepsilon_\pm ^{\parallel} & = & \varepsilon_x^{\parallel}
\pm i\varepsilon_y^{\parallel}=-\cos \theta e^{\pm i\phi }, \nonumber\\
   \varepsilon_z^{\parallel} & = & \sin \theta ,\nonumber \\[-7.5pt]
 \label{eq:polarizdef}\\[-7.5pt]
   \varepsilon _{}^\perp & = & \sin \phi
\hat x-\cos \phi \hat y, \nonumber\\
   \varepsilon _\pm ^\perp & = & \varepsilon_x^\perp
 \pm i\varepsilon_y^\perp =\mp ie^{\pm i\phi },\nonumber\\
   \varepsilon_z^\perp & = & 0 \nonumber
\end{eqnarray}
and the ``vertex'' functions, which in the notation of DH86 have the form
\begin{eqnarray}
   \Lambda _{\ell ,m}(\beta )=(-i)^{G-S}\left( {{{S!} \over {G!}}}
   \right)^{1/2}2^{-(G+S)/2}(\beta ^*)^\ell \beta ^m \left( {{{\left| \beta
   \right|^2} \over 2}} \right)^{-S}L_S^{G-S}\left( {{{\left| \beta
   \right|^2} \over 2}} \right) ,
 \label{eq:vertfunc_def}
\end{eqnarray}
where
\begin{eqnarray}
   G =\max (\ell ,m), && \ S=\min (\ell ,m), \nonumber \\[-5.5pt]
 \label{eq:GSbetadef}\\[-5.5pt]
   \beta = -i{{(k_x+ik_y)} \over {\sqrt B}}, &&
   \ \beta '=i{{(k'_x+ik'_y)} \over {\sqrt B}} \nonumber
\end{eqnarray}
and $L^a_n(x)$ are the associated Laguerre polynomials.  

\section{SCATTERING OF RELATIVISTIC ELECTRONS}

We consider in this study the particular problem of scattering of
photons from relativistic electrons, common to a variety of
astrophysical phenomena.  Such relativistic scattering leads to
considerable simplification of the algebra.  In this section we develop
expressions for scattering of ultra-relativistic electrons with $\gamma
\gg 1$ moving parallel to the magnetic field lines.  Generally, the
photon may have any angle of incidence, $\psi_i$, in the laboratory
frame with respect to the magnetic field.  Due to the large $ \gamma
$'s, the laboratory angle, $\psi_i$, gets Lorentz contracted to
$\theta=\psi_i/2\gamma \sim 0$ degrees in the electron rest frame.  The
magnetic, nonrelativistic Thomson cross section (Canuto, Lodenquai \&
Ruderman 1971, Blandford \& Scharlemann 1976, and Herold 1979) consists
of two parts dependent on the incident photon angle as shown in the
expression
\begin{equation}
   \sigma _{\rm Thomson}=\sigma_{\rm T}\left[ {\sin ^2\theta +{{\omega ^2} \over
   2}(1+\cos ^2\theta )\left( {{1 \over {(\omega -B)^2}}+{1 \over {(\omega
   +B)^2}}} \right)} \right]. 
\end{equation}
Well below the resonance for angles, $\theta \ll \omega /B\ll 1$,
corresponding to $\psi_i\ll 2 \gamma \omega/B$ in the laboratory frame,
the $\sin^2 \theta$ term is smaller than the resonant term.  In the
case of isotropic incident photons in the laboratory frame (where $0 <
\psi_i < \pi$), this constraint requires $\omega \gg \pi B/2\gamma$,
easily achieved with the large $\gamma$ of relativistic electrons
considered throughout this paper.  Under this assumption, the incident
photon is parallel to the magnetic field lines and has no perpendicular
momentum; hence
\begin{eqnarray}
  \theta &=&0, \nonumber \\
  k_\bot &=&0, \nonumber \\
  \varepsilon _z & =&0, \\
  \Phi _1 =-\Phi _2 &=&0,\nonumber \\
  \beta & =&0. \nonumber
\end{eqnarray}
The coordinate system can be arbitrarily set so that the azimuthal
angle, $\phi=0$. Since the incident photon is parallel to the z-axis,
there is no component of the polarization vector, $\varepsilon_z$,
along the z-axis, thereby, eliminating several terms in the $F$'s.  The
vertex functions associated with the incident photon become Kronecker
delta functions, $\Lambda_{l,m}(0)=\delta_{lm}$.  This has the advantage
of eliminating the sum over the intermediate states as only certain
specific states contribute depending on the final Landau state,
$\ell$.  The vertex functions associated with the scattered photon can
be written in terms of a single function using the following recursion
relations
\begin{eqnarray}
\Lambda _{\ell -1,0}(\beta' )&=&i{{\sqrt {2\ell }} \over {\beta'^*}}
\Lambda _{\ell ,0}(\beta' ), \nonumber \\
\Lambda _{\ell ,1}(\beta' )&=&{{i\sqrt 2} \over {\beta'^*}}\left( {\ell
-{{\left| \beta'  \right|^2} \over 2}} \right)\Lambda _{\ell ,0}(\beta' ),
\nonumber \\
\Lambda _{\ell -1,1}(\beta' )&=&-{{2\sqrt \ell } \over {(\beta'^*)^2}}
\left( {\ell -1-{{\left| \beta'  \right|^2} \over 2}}
\right)\Lambda _{\ell ,0}(\beta' ), \\
\Lambda _{\ell +1,0}(\beta' )&=&-i{{\beta'^*} \over {\sqrt {2(\ell
+1)}}}\Lambda _{\ell ,0}(\beta' ), \nonumber \\
\Lambda _{\ell -2,0}(\beta' )&=&-{{2\sqrt {\ell (\ell -1)}} \over
 {(\beta'^*)^2}}\Lambda _{\ell ,0}(\beta' ). \nonumber
\end{eqnarray}
The $F$ terms associated with each Feynman diagram can then be rewritten,
with the common vertex function $\Lambda_{l,0}(\beta')$ being factored
out of the scattering amplitudes to generate coefficients $G_i$ such
that $G_i\Lambda _{\ell ,0}(\beta')=F_{n,i}^{(1)}+F_{n,i}^{(2)}$,
where $i$ = flip and no-flip.  Using the identity
\begin{equation}
   \Bigl| {\Lambda _{\ell ,0}(\beta' )} \Bigr|^2={{ {\left| \beta'
   \right|^{2\ell}} } \over {2^\ell \ell !}}={1 \over {\ell !}}
   \left( {{{\omega'^2\sin ^2\theta'} \over {2B}}} \right)^\ell ,
 \label{eq:lambda_l0}
\end{equation}
the differential cross section can be expressed compactly as
\begin{equation}
   {{d\sigma _{\parallel,\perp }} \over {d\cos \theta '}} = {{3\sigma_{\rm T}}
   \over {16\pi }}{{\omega '^2e^{-\omega '^2\sin ^2\theta '/2B}} \over
   {\omega (2+\omega -\omega ')(\omega '+\omega \omega '(1-\cos \theta')
   -\omega '^2\sin ^2\theta ')}}
   {1 \over {\ell !}}\left( {{{\omega '^2\sin ^2\theta '} \over
   {2B}}} \right)^\ell \; {G_{\parallel,\perp }} ,
 \label{eq:dsig_final}
\end{equation}
where
\begin{equation}
   G_{\parallel} =\hat{G} _{no\,flip}^{\parallel}
   + \hat{G}_{flip}^{\parallel},\quad
   G_\perp = \hat{G} _{no\,flip}^\perp + \hat{G} _{flip}^\perp ,
\end{equation}
and 
%
%
%
\begin{eqnarray}
\hat{G} _{no\,flip}^{||}&=&\int\limits_0^{2\pi } {\left|
{G_{no\,flip}^{||,||}} \right|^2d\phi '}=\int\limits_0^{2\pi } {\left|
{G_{no\,flip}^{\bot ,||}} \right|^2d\phi '}\nonumber \\
&=& 2\pi \left\{ {\,\left[ {(B_1+B_3+B_7)\cos \theta '-(B_2+B_6)\sin
\theta '} \right]^2+\left[ {B_4\cos \theta '-B_5\sin \theta '}
\right]^2} \right\},\nonumber \\
\hat{G} _{no\,flip}^\bot& =&\int\limits_0^{2\pi } {\left|
{G_{no\,flip}^{||,\bot }} \right|^2d\phi '}=\int\limits_0^{2\pi }
{\left| {G_{no\,flip}^{\bot ,\bot }} \right|^2d\phi '}\nonumber \\
&=& 2\pi \left\{ {\,(B_1-B_3-B_7)^2+B_4^2} \right\}, \nonumber\\[-5.5pt]
 \label{eq:Gexpressions}\\[-5.5pt]
\hat{G} _{flip}^{||}&=&\int\limits_0^{2\pi } {\left| {G_{flip}^{||,||}}
\right|^2d\phi '}=\int\limits_0^{2\pi } {\left| {G_{flip}^{\bot ,||}}
\right|^2d\phi '}\nonumber \\
&=& 2\pi \left\{ {\,\left[ {(C_1+C_3+C_7)\cos \theta '-(C_2+C_6)\sin
\theta '} \right]^2+\left[ {C_4\cos \theta '-C_5\sin \theta '}
\right]^2} \right\}, \nonumber \\
\hat{G} _{flip}^\bot &=&\int\limits_0^{2\pi } {\left| {G_{flip}^{||,\bot
}} \right|^2d\phi '}=\int\limits_0^{2\pi } {\left| {G_{flip}^{\bot ,\bot
}} \right|^2d\phi '}\nonumber\\
&=& 2\pi \left\{ {\,(C_1-C_3-C_7)^2+C_4^2}\right\}.\nonumber
\end{eqnarray}
In these developments, the $\phi'$ dependence and the imaginary terms
are isolated in the polarization components and the phase exponentials,
leading to elementary integrations over the azimuthal angle, $\phi'$.
The $B$ and $C$ terms have the following forms:
\begin{eqnarray}
B_1&=&{{\left[ {2\omega -\omega \omega '(1-\cos \theta ')} \right]}
\over {2(\omega -B)}},\nonumber \\
B_2&=&-{{\left[ {(\omega -\omega '\cos \theta ')\left( {2lB-\omega
'^2\sin ^2\theta '} \right)+2lB\omega } \right]} \over {2\omega '\sin
\theta '(\omega -B)}},\nonumber \\
B_3&=&{{lB\left( {2lB-2B-\omega '^2\sin ^2\theta '} \right)} \over
{\omega '^2\sin ^2\theta '(\omega -B)}},\nonumber \\
B_4&=&-{{\left[ {2\omega '+\omega \omega '(1-\cos \theta ')-\omega
'^2\sin ^2\theta '} \right]} \over {2\left[ {\omega \omega '(1-\cos
\theta ')-\omega -B} \right]}},\nonumber \\
B_5&=&-{{(\omega -\omega '\cos \theta ')\omega '\sin \theta '} \over
{2\left[ {\omega \omega '(1-\cos \theta ')-\omega -B}
\right]}},\nonumber \\
B_6&=&{{\ell B\cos \theta '} \over {\sin \theta '\left[ {\omega \omega
'(1-\cos \theta ')-\omega +B} \right]}},\nonumber \\
B_7&=&{{2\ell (\ell -1)B^2} \over {\omega '^2\sin ^2\theta '\left[
{\omega \omega '(1-\cos \theta ')-\omega +B} \right]}},\nonumber \\[-4pt]
 \label{eq:BCterms}\\[-4pt]
C_1&=&\sqrt {2\ell B}{\omega  \over {2(\omega -B)}},\nonumber \\
C_2&=&-\sqrt {2lB}{{\left[ {2\omega +2\omega ^2-\omega \omega '(1-\cos
\theta ')-2lB+\omega '^2\sin ^2\theta '} \right]} \over {2\omega '\sin
\theta '(\omega -B)}},\nonumber \\
C_3&=&\sqrt {2lB}{{\left( {\omega -\omega '\cos \theta '} \right)\left(
{2lB-2B-\omega '^2\sin ^2\theta '} \right)} \over {2\omega '^2\sin
^2\theta '(\omega -B)}},\nonumber \\
C_4&=&-\sqrt {2\ell B}{{\omega '\cos \theta '} \over {2\left[ {\omega
'\omega (1-\cos \theta ')-\omega -B} \right]}},\nonumber \\
C_5&=&\sqrt {2\ell B}{{\omega '\sin \theta '} \over {2\left[ {\omega
'\omega (1-\cos \theta ')-\omega -B} \right]}},\nonumber \\
C_6&=&-\sqrt {2\ell B}{{\left[ {2\omega '+\omega \omega '(1-\cos \theta
')-\omega '^2\sin ^2\theta '} \right]} \over {2\omega '\sin \theta
'\left[ {\omega '\omega (1-\cos \theta ')-\omega +B} \right]}}, \nonumber \\
C_7&=&\sqrt {2\ell B}{{(l-1)B\left( {\omega -\omega '\cos \theta '}
\right)} \over {\omega '^2\sin ^2\theta '\left[ {\omega '\omega (1-\cos
\theta ')-\omega +B} \right]}}. \nonumber
\end{eqnarray}

Since this combination of expressions for the differential cross
section is for the particular case of scattering of relativistic
electrons with an incident photon angle of zero degrees in the electron
rest frame, the resulting scattering rates are simple and do not
possess the sum over Bessel functions as in Bussard et al. (1986).
While there are four different possibilities for the scattering of
polarized photons, ($\parallel \rightarrow \parallel$,  $\perp
\rightarrow \parallel$, $\parallel \rightarrow \perp$, and $\perp
\rightarrow \perp$), for this special case under consideration, the
expressions for the $\parallel \rightarrow \parallel$, $\perp
\rightarrow \parallel$ scattering have the same form, as well as those
for the  $\parallel \rightarrow \perp$, and $\perp \rightarrow \perp$
scattering, as indicated above, resulting in two polarization channels
in which the scattering process will produce either parallel or
perpendicular polarized photons.

Observe also that the differential cross section is dependent on the
final Landau state, $\ell$.  Hence, in order to derive the complete
contribution, a sum must be performed over all the contributing Landau
states.  Since the energy of the scattered photon may be expressed as
\begin{equation}
   \omega '={{2(\omega -\ell B)} \over {1+\omega (1-\cos \theta ')+\left[
   {\left( {1+\omega (1-\cos \theta ')} \right)^2-2(\omega -\ell B)
   \sin^2\theta '} \right]^{1/2}}}\;\; ,
 \label{eq:omega_fin}
\end{equation}
where $\ell$ is the final Landau state of the scattered electron, each
final state has an energy threshold of $\ell B$.  Therefore, the
maximum contributing Landau state quantum number, $\ell_{max}$, to the
cross section is pinned to the photon energy in cyclotron units:
$\omega/B-1<\ell_{max}<\omega/B$.

\section{ANGLE-INTEGRATED CROSS SECTIONS}

The differential cross section can be numerically integrated over
$\theta' $ using a Romberg integration scheme to obtain the energy
dependent cross section.  In Figure 2, we display the QED, exact
angle-integrated cross section (solid curves) for the indicated
magnetic fields, in units of $B_{\rm cr}$, as a function of the incident
photon energy, $\omega/B$, in cyclotron energy units.  We have
averaged over the final spin of the electron and over the polarization
of the scattered photon. For this particular case in the scattering of
relativistic electrons, there is only one resonance occurring at the
fundamental cyclotron resonance of $\omega_B=B$.  We scale the photon
energy by the cyclotron energy so that the resonance occurs at the same
place independent of the magnetic field, $B$.  For comparison, we also
plot in the figures the nonrelativistic Thomson limit (dot-dashed
curves) and the Klein-Nishina (dotted curves) predictions.  The solid
circles are the result of an approximation that is discussed later in
section 6.  For the exact calculation, we have summed over the all the
contributing final Landau states.

Above the resonance, the exact cross section approaches the
Klein-Nishina cross section.  As expected for smaller fields, the
convergence occurs at lower photon energies, as seen in the case of
$B=0.1$.  At this field strength, typical of radio pulsars, there are
no significant deviations from the Thomson limit below the resonance
and from the Klein-Nishina limit above the resonance.  The main
discrepancy occurs right above the resonance where the two limiting
cases do not match the exact cross section.  As the field strength
increases, the exact cross section below the resonance drops
significantly beneath the Thomson limit by over a factor of ten in the
case of $B=100$  For scattering above the resonance at these high
fields, there are deviations between the exact cross section and the
Klein-Nishina cross section.  However, as the energy increases, the
exact cross section and the Klein-Nishina cross section appear to
converge as seen in the cases of $B=0.1$ and $B=1$.

The trend as $B$ increases, evident in Figure 2, is for the magnitude
of the cross section to drop at all energies, while the width of the
resonance increases (for $B\geq 1$, when scaled in units of the
cyclotron energy, this width actually declines). Since the resonance is
formally divergent, the common practice (Xia et al. 1985, Daugherty \&
Harding 1989 and Dermer 1990) is to truncate it at $\omega=B$ by
introducing a finite width $\Gamma$ equal to the cyclotron decay width
(inverse lifetime) for the $\ell = 1 \rightarrow 0 $ transitions,
corresponding to decay of an excited intermediate electron state.  The
procedure is to replace the resonant $(\omega-B)^2$ denominator (e.g.
see equation~(\ref{eq:dsig_leq0_approx}) below) by $\left[(w -B)^2 +
G^2/4\right]$.  In the $B\ll 1$ regime, the cyclotron decay width
assumes the well-known result $\Gamma = 4\alpha B^2/3$ in dimensionless
units.  When $B\gg 1$, Latal (1986) deduced that $\Gamma$ is
proportional to $B^{1/2}$, a dependence that can be inferred from the
exact decay widths enunciated in equation (17) of Herold, Ruder \&
Wunner (1982) by noting that the angular distribution of radiation in
cyclotron transitions is still quasi-isotropic (and not strongly
beamed) for highly-supercritical fields.  When substituted into the
Lorentz profile prescription for truncating the resonance, these widths
lead to the areas under the resonance (i.e. when integrating over
$\omega$) being independent of $B$ in the magnetic Thomson regime of
$B\ll 1$, and scaling as $B^{1/2}$ when $B\gg 1$; these results can be
deduced using the $\ell=0$ approximation derived in equation
(\ref{eq:dsig_leq0_approx}).

This area is an approximate measure of the importance of resonant
Compton scattering for a particular situation.  To see this, note that
astrophysical problems generally have a broad energy distribution of
beamed relativistic electrons interacting with not-so-highly-beamed low
energy (soft) photons, so that a particular electron energy and soft
photon angle (with respect to $B$) determines the value of $\omega$ in
the electron rest frame.  Summing over the electron energies and
incoming photon angles amounts to a weighted integration of the area
under the cross section.  The weighting function is usually not very
sensitive to $\omega$, so that the area under the curves gives a
representative indication of the strength of resonant (as opposed to
non-resonant) scattering provided the resonance is not too broad.
Hence, it can be inferred that the resonant process is more important
in supercritical fields than when $B\ll 1$.  This conclusion stands even
when it is noted that the magnetic Compton resonance is not truly
cyclotronic in nature: the contribution of the $\ell=1$ transition to
the right wing of the resonance plus the spread of parallel momenta
introduce non-Lorentzian distortions to the resonance profile.

The photon polarization-dependent cross sections can also be easily
obtained for the given integration of the $G$ terms shown above.  In
Figure 3, we present the QED Compton scattering cross section as a
function of energy for magnetic fields of 0.1 and 10 times $B_{\rm cr}$.
As mentioned earlier, for this particular case of scattering of
relativistic electrons, there are two polarization scattering channels,
in which the scattering leads to photons with parallel polarization
(dashed curve) and with perpendicular polarization (dotted curve).  The
total cross section is shown as a solid curve.  In the $B=10$ case,
above the resonance, the scattering process preferably produces photons
with parallel polarization, whereas below the resonance, the channel
producing perpendicularly polarized photons dominates.  This behavior,
where perpendicular-polarized scattered photons dominate below the
resonance and parallel-polarized scattered photons dominate above the
resonance, is characteristic of the magnetic-relativistic cross
section.  In the nonrelativistic case, the perpendicular polarization
channel will be three times larger than the parallel polarization
channel, but has the same shape at all photon energies, as observed
from equation~(\ref{eq:dsig_leq0_approx}) letting
$\omega'\rightarrow\omega$ and integrating over $\cos\theta' $.  As can
be seen in the Figure 3 for the $B=0.1$ case, $\parallel$-polarization
in the QED cross section dominates at low fields, thus the switching to
$\perp$-polarization dominance at high fields is a relativistic effect.
In the Klein-Nishina scattering there is no magnetic field and the
initial photon polarization is important in determining the final
photon polarization.  In this case, there are three channels,
$\parallel\rightarrow\parallel$, $\perp\rightarrow\perp$, and two
degenerate switching channels $\perp\rightarrow\parallel$ and
$\parallel\rightarrow\perp$.  In the Klein-Nishina regime, right above
the resonance, the $\perp\rightarrow\perp$ channel dominates over the
$\parallel\rightarrow\parallel$ channel and the switching channels
$\perp\rightarrow\parallel$ and $\parallel\rightarrow\perp$ which have
the smallest contribution.  Far above the resonance at high photon
energies, the cross sections for the three channels merge as in the
magnetic cross section in Figure 3.

In Figure 4, we show the contributions of the indicated final Landau
states to the total cross section for the indicated field strength of
10 times $B_{\rm cr}$.  The total cross section, represented by a
thick-solid curve, is a result of summing over all contributing Landau
states.  Below the resonance, only the $\ell = 0$ final state
contributes (dotted curve)  to the cross section due to the previously
mentioned threshold associated with each $\ell$.  The curve associated
with $\ell = 1$, having a similar shape as the $\ell = 0$ curve, is
plotted also as a dotted curve. The light solid curves represent a
select group of the indicated higher final Landau states.  As the
photon energy increases, higher $\ell$ states may contribute more
significantly than lower ones.  For example, above a photon energy of
50, the $\ell = 10$ state contributes more to the overall cross section
than the $\ell = 0$ or 1 states (dotted curves).  Clearly for
scattering above the resonance, many final states must be included for
computational accuracy.

As mentioned earlier, the cross section of DH86 was derived using
Johnson-Lippmann (JL) electron wavefunctions which do not correctly
describe the spin-dependence of the S-matrix elements, but produce
correct spin-averaged S-matrix elements.  Thus, we have averaged over
the initial and summed over final electron spin states in the
expressions we derived.  However, there is still a small error in the
JL cross section at and above the cyclotron resonance, due to the
spin-dependence of the intermediate states.  This error is negligible
for $B < 0.1$ but grows with $B$ for $B > 0.1$.  We have evaluated this
error through a numerical comparison of the JL cross section of DH86,
derived in this paper that neglect the decay width of the intermediate
states, with the cross section derived by Sina (1996) who used
Sokolov-Ternov (ST) wavefunctions.  For the case of the scattering of
relativistic electrons with $\psi_i = 0$, the two spin-averaged cross
sections agree below the cyclotron energy, $B$, but do not quite
agree for $\omega \geq B$.  We find that for $B = 0.1$, there is a
small error of 0.01\% at $\omega =B$ in the spin-averaged cross section
and for $B = 10$, there is a somewhat larger error of 0.4\% at
$\omega =B$.  The increasing error with increasing magnetic field is due
to the intermediate states having non-zero momentum which increases the
difference between JL and ST cross sections at higher $B$ fields.  Thus
the simplified expressions derived in this paper are accurate in their
regions of validity. Furthermore, the cyclotron energy is high enough
in supercritical fields that scattering above the resonance is less
important than it is in subcritical fields.  We plan to use the ST
cross section in future derivation of simplified expressions for the
scattering cross section above the cyclotron resonance.

\section{SCATTERING TO $\ell=0$ FINAL STATES - BELOW THE RESONANCE}

Due to the presence of the resonance in the cross section, the
scattering process will try to select out resonant scattering, if the
geometry and the kinematics permit.  As the magnetic field increases,
the photon energy, in the electron rest frame, required for resonant
scattering (i.e., $\omega =B$) increases.  If the source of photons is
limited by blackbody temperatures and the field strength is large, the
scattering will predominately occur much below the resonance.  The
cross sections, described here, are in the rest frame of the electron.
If the electron is moving, the Lorentz-transformed photon energy in the
electron rest frame is given by
\begin{equation}
   E_{rest}=E_{lab}\gamma (1-\beta \cos \psi _i),
\end{equation}
where $E_{rest}$ and $E_{lab}$ are the energies of the photon in the
rest and laboratory frames, respectively, and $\psi_i$ is the
laboratory angle of the photon with respect to the electron direction.
For small angles $\psi_i \sim 0$, where the photon is moving in the
same direction as the electron, the photon energy is red shifted,
$E_{rest}\sim E_{lab}/(2\gamma)$.  For large angles $\psi_i\sim \pi$,
where the photon and electron are colliding head on, the photon energy
is blue shifted, $E_{rest}\sim 2\gamma E_{lab}$.  For perpendicular
scattering, the photon energy is also blue shifted, $E_{rest}\sim
\gamma E_{lab}$. In general, when the incident angle, $\psi_i >
\sqrt{2/\gamma}$, the photon energy will be blue shifted. For
relativistic electrons, the geometry of the interaction becomes
important in determining whether the scattering takes place above or
below the resonance.

If the scattering occurs below the resonance, the only contribution to
the cross section is from the $\ell = 0$ final state, since the final
Landau state, $\ell$, state has an energy threshold of $\ell B$.  Using
the following identity valid for $\ell=0$,
\begin{equation}
   \omega '^2\sin ^2\theta '=2\omega '-2\omega 
   -\omega \omega '(1-\cos \theta'),
\end{equation}
in the denominator of equation~(\ref{eq:dsig_final}), the expression for
the exact, QED, cross section of the $\ell = 0$ final state has the
following form
\begin{equation}
   {{d\sigma } \over {d\cos \theta '}}={{3\sigma_{\rm T}} \over {16\pi
   }}{{\omega '^2} \over {\omega (2+\omega -\omega ')(\xi -\omega')}}
   e^{-\omega '^2\sin ^2\theta '/2B}\;  G \quad ,
\end{equation}
where
\begin{equation}
   \xi =2\omega -\omega \omega '(1-\cos \theta ').
\end{equation}
All the $C$ terms associated with the spin flip drop out for the $\ell =
0$ case, in addition the $B_3$, $B_6$ and $B_7$ terms are zero.  The
remaining nonzero $B$ terms simplify to the following
\begin{eqnarray}
B_1&=&{\xi  \over {2(\omega -B)}},\nonumber \\
B_2&=&{\eta  \over {2(\omega -B)}},\nonumber \\
B_4&=&-{{-\xi } \over {2(\omega -\xi -B)}}, \\
B_5&=&-{\eta  \over {2(\omega -\xi -B)}},\nonumber
\end{eqnarray}
where
\begin{equation}
   \eta =(\omega -\omega '\cos \theta ')\omega '\sin \theta '.
\end{equation}
The exact polarization-dependent and averaged cross sections for
$\ell=0$ can be expressed as
\begin{eqnarray}
{{d\sigma ^{||\to ||}} \over {d\cos \theta '}}&=&{{d\sigma ^{\perp \to
||}} \over {d\cos \theta '}}={{3\sigma_{\rm T}} \over {32}}{{\omega
'^2e^{-\omega '^2\sin ^2\theta '/2B}} \over {\omega (2+\omega -\omega
')(\xi -\omega ')}}\nonumber \\[-5.5pt]
 \label{eq:dsig_leq0_exact}\\[-5.5pt]
& & \times\left\{ {\left( {\xi \cos \theta '-\eta \sin \theta '}
\right)^2\left[ {{1 \over {(\omega -B)^2}}+{1 \over {(\omega -\xi
-B)^2}}} \right]} \right\},\nonumber \\ {{d\sigma^{||\to \perp }} \over
{d\cos \theta '}}&=&{{d\sigma ^{\perp \to \perp }} \over {d\cos \theta
'}}={{3\sigma_{\rm T}} \over {32}}{{\omega '^2e^{-\omega '^2\sin ^2\theta
'/2B}} \over {\omega (2+\omega -\omega ')(\xi -\omega ')}}\nonumber \\
& & \times \left\{ {\xi ^2\left[ {{1 \over {(\omega -B)^2}}+{1 \over
{(\omega -\xi -B)^2}}} \right]} \right\}, \\ {{d\sigma
^{ave}} \over {d\cos \theta '}}&=&{{3\sigma_{\rm T}} \over {32}}{{\omega
'^2e^{-\omega '^2\sin ^2\theta '/2B}} \over {\omega (2+\omega -\omega
')(\xi -\omega ')}}\nonumber \\
 & & \times \left\{ {\left[ {\left( {\xi \cos \theta '-\eta \sin
\theta '} \right)^2+\xi ^2} \right]\left[ {{1 \over {(\omega -B)^2}}+{1
\over
{(\omega -\xi -B)^2}}} \right]} \right\}. \nonumber
\end{eqnarray}
The average cross section is obtained by summing over the final and
averaging over the initial photon polarizations.

\section{APPROXIMATING THE $\ell=0$ FINAL STATES}

An approximation to the exact $\ell=0$ differential cross section can
be given by assuming that the scattering is significantly below the
resonance, where $\omega < B$, and also $\omega<1$.  We make
this approximation by keeping only terms to first order in $\omega $
and $ \omega ' $ in the two forms,
\begin{eqnarray}
\xi &=& 2 \omega -\omega \omega '(1-\cos \theta ')\to 2\omega ,\nonumber \\
\eta &=& (\omega -\omega '\cos \theta ')\omega '\sin \theta '\to 0.
\end{eqnarray}
This leads to the following approximate expressions
\begin{eqnarray}
{{d\sigma ^{||\to ||}} \over {d\cos \theta '}}&\approx& {{d\sigma ^{\perp
\to ||}}
\over {d\cos \theta '}}={{3\sigma_{\rm T}} \over 8}{{\omega \omega '^2\cos
^2\theta '}
\over {(2\omega -\omega ')}}\left[ {{1 \over {(\omega -B)^2}}+{1 \over
{(\omega
+B)^2}}} \right],\nonumber \\
{{d\sigma ^{||\to \perp }} \over {d\cos \theta '}}&\approx&
{{d\sigma ^{\perp \to \perp }} \over {d\cos \theta '}}={{3\sigma_{\rm T}} \over
8}{{\omega \omega '^2} \over {(2\omega -\omega ')}}\left[ {{1 \over {(\omega
-B)^2}}+{1 \over {(\omega +B)^2}}} \right], \nonumber\\[-5.5pt]
 \label{eq:dsig_leq0_approx}\\[-5.5pt]
{{d\sigma ^{ave}} \over
{d\cos \theta '}}&\approx& {{3\sigma_{\rm T}} \over 8}{{\omega \omega '^2(1+\cos
^2\theta ')} \over {(2\omega -\omega ')}}\left[ {{1 \over {(\omega
-B)^2}}+{1
\over {(\omega +B)^2}}} \right]. \nonumber
\end{eqnarray}
The nonrelativistic approximation would further lead to $\omega'
=\omega$, resulting in the nonrelativistic expressions found in Herold
(1979).

These expressions, while much simpler than the previous ones, are still
complex due to the functional form of $\omega'$, having an angular
dependence in its square root.  Yet they are manageable and can be
integrated over $\cos\theta'$.  After careful algebra, integration over
$\cos\theta'$ yields the following polarization-dependent and averaged,
approximate cross sections
\begin{eqnarray}
\sigma ^{||\to ||}&=&\sigma ^{\perp \to ||}={{3\sigma_{\rm T}} \over {16}}\left\{
{g(\omega )-h(\omega )} \right\}\left[ {{1 \over {(\omega -B)^2}}+{1 \over
{(\omega +B)^2}}} \right],\nonumber \\
\sigma ^{||\to \perp }&=&\sigma ^{\perp \to \perp
}={{3\sigma_{\rm T}} \over {16}}\left\{ {f(\omega )-2\omega h(\omega )}
\right\}\left[
{{1 \over {(\omega -B)^2}}+{1 \over {(\omega +B)^2}}} \right], \\
\sigma^{ave}&=&{{3\sigma_{\rm T}} \over {16}}\left\{ {g(\omega )+f(\omega
)-(1+2\omega
)h(\omega )} \right\}\left[ {{1 \over {(\omega -B)^2}}+{1 \over {(\omega
+B)^2}}}
\right].\nonumber
\end{eqnarray}
where
\begin{eqnarray}
g(\omega )&=&{{\omega^2 (3+2\omega )+2\omega} \over {\sqrt {\omega
(2+\omega )}}}\ln \left( {1+\omega -\sqrt {\omega (2+\omega )}}
\right)+{\omega  \over 2}\ln (1+4\omega )\nonumber \\ & & +\omega
(1+2\omega )\ln (1+2\omega )+2\omega ,\nonumber \\ f(\omega )&=&-\omega
^2\ln (1+4\omega )+\omega (1+2\omega )\ln (1+2\omega ), \\
h(\omega )&=&\left\{
\begin{array}{ll}
{{\omega ^2} \over {\sqrt {\omega (2-\omega )}}}\tan ^{-1}\left(
{{{\sqrt {\omega (2-\omega )}} \over {1+\omega }}} \right), & \mbox{for
$\omega <2$,} \\ {{\omega ^2} \over {2\sqrt {\omega (\omega
-2)}}}\ln \left( {{{\left( {1+\omega +\sqrt {\omega (\omega -2)}}
\right)^2} \over {1+4\omega }}} \right), & \mbox{for $\omega >2$.}
\end{array}
\right. \nonumber
\end{eqnarray}
While one might expect $h(\omega)$ to become imaginary when $\omega>2$,
given the $(2-\omega)$ factor in the square-root terms in first
expression for $h(\omega)$, the expression is completely general and
remains real even when $\omega>2$.  For the purpose of numerical
calculations in a computer code, we introduce this second expression
for $h(\omega)$ that can be coded using a natural logarithm when
$\omega>2$.

Back in Figure 2, the solid circles represent the polarization-averaged
cross section obtained from the above analytical approximation.  The
approximation is valid in the region below $\omega<1$ corresponding to
$\omega/B<1/B$ along the photon energy axis in Figure 2.  In
this region, it agrees very well with the exact $\ell=0$ cross
section.  Above the region of validity, the approximation over
estimates the exact $\ell=0$ cross section.  However, the approximation
does surprisingly well, when compared to the exact cross section,
extrapolating above the region of validity. While the analytical
approximation is a result of integrating the approximation to the exact
$\ell=0$ differential cross section, it remains close to the total
cross section at high $\omega$ above the resonance $(\omega >B)$
even for high field strengths.

\section{ANGULAR DISTRIBUTIONS}

We present in Figure 5, the differential cross section, $ d \sigma / d
\theta' $, for a magnetic field of $B=10 B_{\rm cr}$.  In order to
understand best the behavior of the angular distributions, we have
plotted the differential cross section as a function of the logarithm
of the scattered angle, $\theta'$, of the photon in the electron rest
frame.  We sample the angular distributions beginning below the
resonance, where only the $\ell=0$ final Landau state contributes, to
high above the resonance where many Landau states contribute up to the
threshold $\ell_{max}$: $\omega/B-1 < \ell_{max} < \omega/B$.  The
incident photon energies for each panel are indicated in units of the
cyclotron energy, $\omega_B=B$.  Also indicated are the final Landau
states of the calculated angular distributions.  As expected, the
$\ell=0$ contribution is strong for all photon energies.  As the photon
energy increases high above the resonance (right panels of the figure),
the angular distribution of the $\ell=0$ state reveals a dip at an
energy-dependent angle in the forward direction.  This dip is very
steep, as indicated by the number of decades the distribution drops
before approaching its minimum.  Care must be exercised in this region,
as one integrates the angular distributions.  As the final Landau state
increases, the angular distributions become more gaussian shaped,
peaking at the same angle for a given photon energy as the minimum in
the $\ell=0$ state.  Above the resonance, the angular distributions
evolve smoothly from a sharp minimum at low Landau states to a maximum
at higher Landau states.  Both the minima and maxima occur at
invariably the same angle, $\theta_o$.  This behavior is due to the
functional form of the following factor in the differential cross
section
\begin{equation}
   f(\omega ,\theta ')={1 \over {\ell!}}\left( {{{\omega '^2\sin ^2\theta '}
   \over {2B}}} \right)^\ell
   \exp\Biggl\{ {-{{\omega '^2\sin ^2\theta '} \over {2B}}} \Biggr\},
\end{equation}
which controls the angular dependence of the differential cross
section, $d\sigma/d\cos\theta'$.  The first part of this function
arises from the vertex functions mentioned earlier in equation
(\ref{eq:lambda_l0}).  The first derivative with respect to $\theta'$
of this function goes to zero an angle, $\theta_o$, given by
\begin{equation}
   \theta_o=\tan ^{-1}\left( {{{\sqrt {1+2\omega }} \over \omega }} \right),
\end{equation}
independent of $\ell$.  Since $f(\omega,\theta')$ is only part of the
overall differential cross section, this expression for $\theta_o$ is
approximate.  Yet, it predicts very well the angle at which the angular
distributions experience minima and maxima with increasing Landau
states.  The scattered photon energy, $\omega_o'$, at which this peak
in the angular distribution occurs is given by
\begin{equation}
   \omega'_o={{1+\omega } \over {1+2\omega }}\left( {1+2\omega -\sqrt
   {(1+2\omega )(1+2\ell B)}} \right).
 \label{eq:omega_op}
\end{equation}
The Landau state, $\ell_s$, at which the angular distribution evolves
from a minimum to a maximum at a given photon energy, $\omega$, occurs
when the second derivative of the above function, $f(\omega,\theta')$,
with respect to $\theta'$ is equal to zero and is given by the
expression
\begin{equation}
   \ell _s={{\omega ^2} \over {2B(1+2\omega )}}.
\end{equation}
The energy of scattered photon at $\ell_s$ and at $\theta_o$ has the
simple form
\begin{equation}
   \omega'_s={{(1+\omega )\omega } \over {1+2\omega }}.
 \label{eq:omega_sp}
\end{equation}
These expressions have served to guide the design of the algorithm that
numerically integrates the angular distributions to obtain the
integrated cross sections shown in Figure 2.  For Landau states below,
$\ell_s$, the angular distributions manifest the steep drop at
$\theta_o$, therefore we integrate the angular distribution in two
parts from $\theta=0$ to $\theta_o$ and from $\theta_o$ to $\pi$.  When
the Landau state is above, $\ell$s, the angular distributions are
gaussian-shaped, and we integrate from $\theta=0$ to $\pi$.

The peak in the total angular distribution will also occur very near
this angle $\theta_o$ as seen in Figure 6.  Here we present the total
angular distributions summed over all the contributing final Landau
states represented by the solid curves. The dashed curves show the
angular distribution for the $\ell=0$ state.  The approximation to the
exact $\ell=0$ differential cross section given in equation
(\ref{eq:dsig_leq0_approx}) (averaged) is plotted as dot-dashed
curves.  Comparing the exact-summed angular distributions (solid
curves) to the approximate distributions (dot-dashed) in Figure 6, one
can see in Figure 2 that the approximation falls below the integrated
cross sections because there is a deficiency in the approximation at
large angles.  As expected, the $\ell=0$ state is the dominant
contribution to the angular distribution at photon energies below and
right above the resonance.  As the photon energy increases well beyond
the resonance, the $\ell=0$ state contributes less significantly and
higher Landau states become increasingly important contributors.
Although the strength of large Landau states decreases rapidly as shown
in Figure 5, they are numerous and contribute significantly when summed
to obtain the overall angular distributions shown in Figure 6.  The
contribution of these higher Landau states occurs near $\theta_o$,
where the total angular distributions peak.  It is at $\theta_o$ where
minima occur in the angular distributions of final states of
$\ell<\ell_s$.  Yet the approximation to the $\ell=0$ final Landau
state peaks at approximately the angle $\theta_o$.  The angular
distributions of the exact, the Klein-Nishina, and the $\ell=0$
approximate cross sections all peak where $\omega'=\omega\ ({\rm
at\ }\theta=\theta_o )$.  This is a result of the fact that the
approximate angular distribution is governed by the kinematics.  The
scattered angle, $\theta'$, is small at $\theta_o$, and the term
$2(\omega-\ell B)\sin^2\theta'$ in the denominator of equation
(\ref{eq:omega_fin}) is also small resulting in an expression very
similar to the Klein-Nishina kinematics.  At scattering angles below
$\theta_o$, there is significant agreement between the exact and
Klein-Nishina angular distributions.  The small-angle, low recoil
scatterings, where one would expect all to agree because of similar
kinematics, does not probe the effects of the field.

Differences are seen at large angle, large recoil scattering, where the
geometry of the magnetic scattering is impacted by Landau excitations.
High above the resonance, magnetic effects become manifested in the
backward direction. Comparisons with the Klein-Nishina angular
distributions for large photon energies, suggest that at backward
angles the Klein-Nishina cross section over estimates the exact,
summed, angular distributions beyond an angle of about 30 degrees.  At
these backward angles, Landau states larger than zero contribute
significantly.  While the term $2(\omega-\ell B)\sin^2\theta'$ might be
small at these angles, the $\ell B$ term in the numerator of $\omega'$
in equation (\ref{eq:omega_fin}) becomes more important for larger
$\ell$'s, and $\omega'$ is significantly less than $\omega$, while
$\omega'$ in the Klein-Nishina kinematics remains close to $\omega$.

A quantity of interest in polar cap cascade models of highly magnetized
gamma-ray pulsars is the mean value of the product $\omega'\sin\theta'$
achieved in resonant Compton scatterings.  This product is a Lorentz
invariant in transformations along the field lines, and represents the
photon energy in the frame of reference where the upscattered photons
move orthogonally to the local field.  Hence, in conjunction with the
value of $B$, this product principally controls the strength of
strong-field photon attenuation processes such as pair creation
$\gamma\rightarrow e^+e^-$ and photon splitting
$\gamma\rightarrow\gamma\gamma$ .  The mean value of
$\omega'\sin\theta'$ in scatterings is therefore extremely informative
for pulsar cascade modelers, and accordingly is plotted as a function
of incident photon energy for different $B$ in Figure 7.  The average
was formed by weighting the differential cross sections such as those
in Figure 5 with $\omega'\sin\theta'$ using equation
(\ref{eq:omega_fin}), summing over quantum numbers $\ell$ and
integrating over $\theta'$, and then dividing by the total cross
section (see Figure 2).  The resulting curves exhibit a generally
increasing function of $\omega$, with structure at the cyclotron
resonances that becomes prominent in critical and supercritical fields
due to the enhanced importance of $\ell >0$ (excited final state)
scatterings above the fundamental; for a given $\theta'$, higher $\ell$
values imply lower final photon energies $\omega'$ (see equation
[\ref{eq:omega_fin}]).  The most salient property of Figure 7 is that
the criterion for the scattered photons to generally rise above pair
threshold is largely insensitive to the value of $B$, and is contingent
upon the initial photon energy exceeding about 5-10 MeV in the electron
rest frame.

Other general properties of $\left\langle
\omega'\sin\theta'\right\rangle$ in Figure 7 can be understood as
follows. Obviously there is naturally no expectation that the behavior
of Figure 7 at low $B$ should mimic the non-magnetic scattering average
at or below the cyclotron fundamental, because the total cross section
does not approach the field-free Thomson limit when $\omega \ll B$ (see
Figure 2).  In fact, contrary to such intuition, at energies well below
the cyclotron fundamental, $\left\langle
\omega'\sin\theta'\right\rangle$  asymptotically approaches
$15\pi\omega/64$ independent of $B$, derivable from the first line of
equation (\ref{eq:dsig_leq0_approx}), a limit identical to the
non-magnetic Thomson average.  This ensues since, while the magnetic
cross section has a different magnitude from the Thomson one, it
possesses the same angular dependence (e.g. see equation (7.1b) of
Rybicki \& Lightman 1979), and $\omega'\rightarrow\omega$ in this
limit, independent of $B$.  Departures from this low energy asymptote
arise when $B \sim 1$.  The analysis of the $\omega \gg B$ case is less
trivial.  Since the Klein-Nishina cross section is reproduced in
sub-critical fields (e.g. see Figure 2), it might be anticipated that
the Klein-Nishina $\left\langle \omega'\sin\theta'\right\rangle_{\rm
KN}$ might result.  This is realized, more or less, with the $B=0.1$
curve in Figure 7, which, if extrapolated, asymptotically approximates
the non-magnetic Klein-Nishina average of
\begin{equation}
   \left\langle {\omega '\sin \theta '} \right\rangle_{\rm KN}
   \mathrel{\mathop{\kern0pt\longrightarrow}
   \limits_{\omega \gg B}}{{9\pi} \over 8}
   {{\sqrt {2\omega }} \over {2\ln (2\omega )+1}}
\end{equation}
when $\omega \gg 10^3$.  For higher $B$, deviations from pure
Klein-Nishina behavior are more apparent, though the approximately
$\omega^{1/2}$ dependence of $\left\langle
\omega'\sin\theta'\right\rangle$ is generally maintained.  The peak
contribution to the cross section arises at scattered angles
$\theta'\approx 1/\omega^{1/2}$ when $\omega \gg 1$, as can be seen
from equation (\ref{eq:omega_op}), a consequence of kinematic
constraints imposed by equation (\ref{eq:omega_fin}).  In fact, this
dependence of $\theta'$ is similar to that for non-magnetic
Klein-Nishina scattering, which possess similar (though not identical)
kinematic restrictions.  Furthermore, $\left\langle \theta'\right\rangle$
is proportional to $\omega$ in this regime (for example, see equations
(\ref{eq:omega_op}) and (\ref{eq:omega_sp})), a dependence borne out by
the Klein-Nishina cross section, though the constants of
proportionality differ, and indeed are a weakly increasing function of
$B$ in the magnetic case here.  Hence, in summary, the quantum
kinematics of magnetic Compton scattering are qualitatively similar to
those of Klein-Nishina scattering, and control the behavior of
$\left\langle \omega'\sin\theta'\right\rangle$ when $\omega \gg B$ and
concomitant deviations from the non-magnetic case.

\section{DISCUSSION}

In this paper, we have extended the work of DH86 by exploring the
regime of super-critical fields and by providing simplified and
explicitly real expressions for the exact differential cross section
for Compton scattering in the presence of strong magnetic fields.  We
have derived simple analytic approximations for both the differential
and total cross sections, in the special case of scattering by highly
relativistic electrons, important to a variety of astrophysical
sources.  These results will be very useful in studying the effects of
Compton scattering in the ultra-strong magnetic fields believed to be
present near stellar surfaces of SGRs and AXPs.   They also provide
much more accurate expressions for modeling Compton scattering in the
fields, $B > 0.1$, of many pulsars.  From the comparison of the exact,
angle-integrated cross section with the limiting cases of the
non-relativistic Thomson and the Klein-Nishina cross sections (used in
neutron star applications throughout the literature) depicted in Figure
2, we can draw the following conclusions about scattering in increasing
high fields:  (i) below the resonance, the exact cross section is
depressed below the magnetic Thomson cross section (when $\omega \geq
m_ec^2$) differing by an order of magnitude or more for fields $B>10$;
(ii) at the resonance, the exact cross section is dramatically reduced
below the Thomson cross section, but the width of the resonance
increases; (iii) far above the resonance, the exact cross section
approaches the Klein-Nishina cross section, as expected for large
photon energies, with the energy where the two merge an increasing
function of $B$.  The overall effect of the strong fields is to lower
the cross section at all incident photon energies $\omega\geq m_ec^2$,
decreasing the electron scattering opacity.

Focusing on large photon energies above the resonance, the magnetic
field has a smaller perturbation on the interaction than around the
resonance and below, and the exact cross section tends toward the
Klein-Nishina limit, a feature that is seen in Figure 2 for $B\leq
10$.  However, the analytical demonstration that these two cross
sections approach each other in this high energy domain requires an
approximation to the sum over many Landau states of the scattered
electron and will be treated in a later paper.  By inspection of the
angular distributions in Figure 6, one can see that the disagreement
between the exact cross section and the Klein-Nishina cross section
occurs for moderate to extreme backward scattering, where the
interaction probes the dominant effects of the magnetic field.  It is
also important to note that as the photon energy increases, the number
of final Landau states contributing to the overall cross section
increases dramatically, as seen in Figure 4, and as expected from the
$\ell< \omega/B$ kinematic constraint.  Care must be exercised in
adding the contributions, as the numerical value of the cross section
becomes very small for large Landau states.

From Figure 2 it is clear that even in relatively low fields $(B
\approx 0.1)$, neither the magnetic Thompson nor the Klein-Nishina
cross sections provides an adequate approximation to the exact cross
section in a region right above the resonance, making a better
representation of the exact cross section important for astrophysical
models.  Since the $\ell=0$ scattering provides the sole contribution
to the cross section below the resonance, we were motivated to use it
as a basis for developing an analytic approximation that is strictly
only valid for $\omega\ll 1$ and $B\ll 1$.  However, satisfyingly, the
analytic approximation to the $\ell=0$ angle-integrated cross section
can be extrapolated to $\omega\gg 1$ domains, and is able to represent
the exact cross section quite well even at super-critical fields $(B
\leq 10)$.  This approximation indeed provides a smooth match to the
exact cross section below the resonance, through the resonance and then
above it where the Klein-Nishina-like reductions take over.

Use of the more accurate numerical results or approximate expressions
that we have given here for resonant scattering will generally diminish
the effects attributed to resonant Compton scattering in astrophysical
models that use a non-relativistic treatment.  This is because the
non-relativistic cross section over-estimates the exact cross section
when extrapolated to the domains (i.e. $\omega \geq 1,\ B\geq 1$) where
relativistic effects are important or critical.  Thus, we expect that
the conditions for scattering to be significant in polar cap
acceleration models will be more restricted: i.e., somewhat higher
magnetic field strengths and soft photon densities will be required,
than previously asserted, for Compton scattering to dominate over
curvature radiation in the energy loss and pair production by primary
particles (Sturner 1995, Harding \& Muslimov 1998).

The cross section for Compton scattering is highly dependent on the
photon polarization.  Incident photons with either parallel or
perpendicular polarization may switch polarization modes in a
scattering.  The calculations of this paper, as seen in Figure 3, show
that the polarization-switching properties below the resonance are the
same for the non-relativistic $(B\leq 1)$ and relativistic cases (i.e.
more photons are scattered into the perpendicular mode).  However for
$B\geq 1$, the polarization-mode switching reverses above the
resonance, so that scattering to the parallel mode is dominant. Such
behavior contrasts that of the non-relativistic limit, where photons of
perpendicular polarization are predominantly produced at all energies.
In the case of the Klein-Nishina scattering where there is no magnetic
field, the initial polarizations do matter in determining the final
polarization resulting in three polarization channels, described
earlier.  Here too, the dominance of photons scattered to perpendicular
polarization is manifested in the relativistic, non-magnetic
Klein-Nishina case.

The polarization dependence of the scattering cross section in high
fields will have significant implications for other
polarization-dependent mechanisms such as pair production and,
especially, photon splitting.  As Baring \& Harding (1998) noted,
photon splitting could dominate pair production at supra-critical
magnetic fields, thereby suppressing pair creation and possibly
accounting for the radio quiescence of SGRs and AXPs.  However,
kinematic selection rules (Adler 1971; Shabad 1975) allow only one
splitting mode to operate in the limit of weak dispersion, that in
which photons with perpendicular polarization $(\perp)$ split into two
photons, each with parallel polarization $(\parallel)$.  Under such
restrictions, photon splitting would occur only once, and then pair
production would take over as the dominant attenuating mechanism.  It
is possible that the dispersion characteristics of the ultra-strong
field environment, or perhaps plasma properties present during outburst
mode of SGRs, may permit the two other splitting modes allowed by CP
(charge-parity) invariance to operate, providing parallel-mode photons
the opportunity to split.  Nevertheless, even if other modes do not
become operational in high fields, Compton scattering below the
resonance is able to convert the photons with parallel polarization
into perpendicular polarization, refueling photon-splitting cascades.
Optical depths for such scattering could be quite significant in SGRs
during their high luminosity gamma-ray outbursts, provided that the
photon field does not dominate the SGR energetics.

While the magnitude of the resonant cross section declines with
increasing $B$, its width increases.  For sub-critical fields, the two
trends compensate each other to produce an area under the resonance
curve (i.e. in Figure 2) insensitive to $B$. For supercritical fields,
the area scales as $B^{1/2}$, as we have noted in Section 4.   This
area is an approximate measure of the importance of resonant Compton
scattering for a particular situation.  Hence it follows that, for a
given soft photon field and electron population, resonant Compton
scattering becomes more significant in magnetar-type fields.  However,
it also becomes more difficult to have photons with energies near the
resonance.

Another quantity of interest to astrophysical modelers is the
expectation of $\omega'\sin\theta'$ in scatterings.  This is because
this quantity, a Lorentz invariant in transformations along the field,
represents the major controlling parameter (apart from $B$) for
determining the rates of photon absorption processes such as pair
creation and photon splitting in strong magnetic fields.  When $ B \geq
5 \times 10^{12} G$, resonant Compton upscattering (i.e. the magnetic
inverse Compton process) can be a major contributor to the gamma-ray
emission of pulsars.  Whether the Compton-upscattered photons produced
by primary electrons can generate pairs, and therefore initiate pair
cascades with steeper synchrotron radiation components, is contingent
upon $ < \omega' \sin\theta' > $ exceeding $2m_ec^2$.  Figure 7 reveals
that this criterion is roughly independent of $B$ for $0.1<B<100$, and
that the necessary condition to spawn cascading is $\omega \geq 5-10\
MeV$ in the  electron rest frame.  This translates into a particular
electron energy and soft photon energy and angle with respect to $B$ in
the observer's frame that is readily identifiable for models.  If the
soft photons are quasi-isotropic thermal X-rays from the surface, then
the primary electron Lorentz factors need to exceed $\gamma \approx
10^5\ (T/10^6 K)^{-1}$ in order to satisfy this pair production
criterion, independent of $B$.  For attenuation by photon splitting,
when $B \geq 0.3$ G, the energy at which splitting optical depths
exceed unity can be below pair creation threshold by a decade or more
(e.g. Baring 1991), so that the required value of
$<\omega'\sin\theta'>$ can be much lower.  Hence the target photon
energies required for the resonant process to seed the splitting
mechanism need only be around $\omega\approx 10\ keV\ -\ 1\ MeV$ for
$B\geq 0.3$ G (and approximately inversely proportional to $B$ in
this instance), a much less stringent requirement than that for pair
cascade initiation.

In conclusion, this paper has provided computations of resonant Compton
scattering in a broad range of sub-critical and super-critical fields
in the particular (but widely applicable) case of ultra-relativistic
electrons moving along a uniform magnetic field.  In doing so, we have
simplified the exact QED differential cross section obtained in this
specialization, and derived a compact analytic expression that
approximates the cross section quite well at energies both below and
above the resonance at the cyclotron fundamental, for fields $B\leq 10$.  
Such an approximation should prove extremely useful to
astrophysical modelers interested in highly magnetized (normal and
anomalous X-ray) pulsars and soft gamma repeaters.  Significant
deviations from the differential Klein-Nishina cross section were found
for large scattering angles, though the total magnetic cross section
was observed to approach the classic Klein-Nishina behavior at energies
well above the resonance.  Polarization properties of resonant
scattering were also explored in detail, revealing that, as with the
magnetic Thomson case, the scattered photons are predominantly of
perpendicular polarization below the resonance.  While this property
persists above the resonance for sub-critical fields, a
polarization-reversal arises in super-critical fields at $\omega >
B$ so that parallel photons dominate the scattered photon
population.  Comprehension of such properties may be critical to model
predictions of the emission from magnetars.

\acknowledgments

We thank Ramin Sina for the use of his computer code that numerically
calculates exact QED Compton scattering with Sokolov \& Ternov electron
spin states.  We would like to express our sincere appreciation for
the generous support of NASA under the Summer Faculty Fellowship
Program, of the Michigan Space Grant Consortium, of the Research
Corporation, and of the NSF under the REU program and through the grant
NSF-9876670.

\appendix
\section{Appendix}
 \label{sec:appendix}

Here the $F^{(j)}_n$ terms that contribute to the scattering amplitudes
that appear in the general expression for differential cross section in
Eq.~(\ref{difforig}) are listed, having been derived in the literature
before (e.g. see DH86).   Each of the $F$ terms consists of two parts
due to the spin-degeneracy (above the ground state) of the final Landau
states of the electron.  For the first diagram in Figure 1, the 
electron spin no-flip and spin flip forms are given by
\begin{eqnarray}
\lefteqn{  F_{n,\,no\,flip}^{(1)} = {1 \over {(2\omega +\omega ^2\sin
^2\theta -2nB)}} }\nonumber\\
 & & \times \left\{
\begin{array}{l}
\left[ \omega (2+\omega -\omega ')+\omega \cos \theta (\omega \cos
\theta -\omega '\cos \theta') \right]\Lambda _{\ell ,n}(\beta ') \Lambda
_{n,0}(\beta )\varepsilon_z\varepsilon '^*_z  \vsp\\ +\left[ \omega
(2+\omega -\omega ')-\omega \cos \theta (\omega \cos \theta -\omega
'\cos \theta ') \right]\Lambda _{\ell,n-1} (\beta ')\Lambda
_{n-1,0}(\beta )\varepsilon _-\varepsilon '^*_+ \nonumber  \vsp\\
 + \sqrt {2nB}(\omega\cos \theta -\omega '\cos \theta ') \left[ {\Lambda
_{\ell,n} (\beta ')\Lambda _{n-1,0}(\beta )\varepsilon _-\varepsilon
'^*_z} +\Lambda_{\ell ,n-1}(\beta ')\Lambda _{n,0}(\beta )\varepsilon
_z\varepsilon '^*_+ \right]  \vsp\\ +\sqrt {2\ell B}\left\{
\begin{array}{l} 
\omega \cos \theta \left[\Lambda _{\ell -1,n}(\beta ')\Lambda
_{n,0}(\beta )\varepsilon_z\varepsilon'^*_-+ \Lambda _{\ell
-1,n-1}(\beta ')\Lambda _{n-1,0}(\beta )\varepsilon_-\varepsilon '^*_z
\right]  \\ +\sqrt {2nB}\left[ {\Lambda _{\ell-1,n}(\beta ')\Lambda
_{n-1,0}(\beta )\varepsilon _-\varepsilon '^*_-- \Lambda_{\ell
-1,n-1}(\beta ')\Lambda _{n,0}(\beta )\varepsilon _z\varepsilon '^*_z}
\right]  \\
\end{array}
\right\} 
\end{array}
\right\}  \nonumber\\[-5.5pt]
 \label{eq:F1def}\\[-5.5pt]
\lefteqn{  F_{n,\,flip}^{(1)} ={1 \over {(2\omega +\omega ^2\sin
^2\theta -2nB)}}  }\nonumber\\
 & &  \times \left\{
\begin{array}{l}
\left[ \omega (2+\omega -\omega')+\omega \cos \theta (\omega \cos \theta
-\omega '\cos \theta ') \right]\Lambda_{\ell -1,n-1}(\beta ')\Lambda
_{n-1,0}(\beta )\varepsilon _-\varepsilon'^*_z  \vsp\\ -\left[ {\omega
(2+\omega -\omega ')-\omega \cos \theta (\omega\cos \theta -\omega '\cos
\theta ')} \right]\Lambda _{\ell -1,n}(\beta ')\Lambda_{n,0}(\beta
)\varepsilon _z \varepsilon '^*_- \vsp\\
 +\sqrt {2nB}(\omega \cos \theta
-\omega '\cos \theta ')  \left[ {\Lambda _{\ell
-1,n}(\beta ')\Lambda _{n-1,0}(\beta )\varepsilon _-\varepsilon
'^*_--\Lambda _{\ell -1,n-1}(\beta ')\Lambda _{n,0}(\beta ) \varepsilon
_z\varepsilon '^*_z} \right]  \nonumber \vsp\\
 -\sqrt{2\ell B}\left\{
\begin{array}{l}
\omega \cos \theta \left[ \Lambda _{\ell ,n}(\beta')\Lambda _{n,0}
(\beta )\varepsilon _z\varepsilon '^*_z-\Lambda _{\ell,n-1}(\beta
')\Lambda _{n-1,0}(\beta ) \varepsilon _-\varepsilon '^*_+\right] \\
+\sqrt {2nB}\left[ \Lambda _{\ell ,n}(\beta ')\Lambda_{n-1,0}(\beta
)\varepsilon _- \varepsilon '^*_z+\Lambda _{\ell ,n-1}(\beta')\Lambda
_{n,0}(\beta )\varepsilon _z \varepsilon '^*_+ \right]
\end{array}
\right\}
\end{array}
\right\} , \nonumber
\end{eqnarray}
where the $\varepsilon$ represents the photon polarization components
defining two orthogonal linear polarization vectors, as defined
in Eq.~(\ref{eq:polarizdef}), and the functions
$\Lambda _{\ell ,m}(\beta )$ are defined in Eq.~(\ref{eq:vertfunc_def})
in terms of associated Laguerre polynomials.  The $F$ terms for the
second (exchange) Feynman diagram in Figure~1 are
\begin{eqnarray}
\lefteqn{  F_{n,\,no\,flip}^{(2)} = {1 \over {(\omega '^2\sin ^2\theta
'-2\omega '-2nB)}}  }\nonumber \\ & &\times \left\{ \begin{array}{l}
-\left[ {\omega '(2+\omega -\omega ')+\omega '\cos \theta '(\omega \cos
\theta -\omega '\cos \theta ')} \right]\Lambda _{\ell ,n}(\beta )\Lambda
_{n,0}(\beta ')\varepsilon_z \varepsilon '^*_z   \vsp\\
 -\left[ {\omega
'(2+\omega -\omega ')-\omega '\cos \theta '(\omega \cos \theta -\omega
'\cos \theta ')} \right]\Lambda _{\ell,n-1} (\beta )\Lambda
_{n-1,0}(\beta ')\varepsilon _+\varepsilon '^*_-  \vsp\\
 +\sqrt {2nB}(\omega\cos \theta -\omega '\cos \theta ') \left[ {\Lambda
_{\ell ,n} (\beta )\Lambda _{n-1,0}(\beta ')\varepsilon _z\varepsilon
'^*_-+\Lambda_{\ell ,n-1} (\beta )\Lambda _{n,0}(\beta ')\varepsilon
_+\varepsilon '^*_z} \right] \nonumber\vsp\\
 +\sqrt {2\ell B}\left\{
\begin{array}{l} 
-\omega '\cos \theta '\left[ {\Lambda _{\ell -1,n}(\beta )\Lambda
_{n,0}(\beta ') \varepsilon _-\varepsilon '^*_z+\Lambda _{\ell
-1,n-1}(\beta )\Lambda _{n-1,0}(\beta ') \varepsilon _z \varepsilon
'^*_-} \right] \\ +\sqrt {2nB}\left[ {\Lambda _{\ell -1,n}(\beta
)\Lambda _{n-1,0}(\beta ')\varepsilon _- \varepsilon '^*_--\Lambda_{\ell
-1,n-1}(\beta )\Lambda _{n,0}(\beta ')\varepsilon _z \varepsilon '^*_z}
\right] \\
\end{array}
\right\} 
\end{array}
\right\} ,\nonumber\\[-5.5pt]
 \label{eq:F2def}\\[-5.5pt]
 \lefteqn{  F_{n,\,flip}^{(2)} ={1 \over {(\omega
'^2\sin ^2\theta '-2\omega '-2nB)}}  }\nonumber\\
 & &\times \left\{
\begin{array}{l}
-\left[ {\omega '(2+\omega-\omega ')+\omega '\cos \theta '(\omega \cos
\theta -\omega '\cos \theta ')} \right]\Lambda _{\ell -1,n-1}(\beta
)\Lambda _{n-1,0}(\beta ')\varepsilon_z\varepsilon '^*_-  \vsp\\ +\left[
{\omega '(2+\omega -\omega ')-\omega '\cos \theta '(\omega \cos \theta
-\omega '\cos \theta ')} \right]\Lambda _{\ell-1,n}(\beta ) \Lambda
_{n,0}(\beta ')\varepsilon _-\varepsilon '^*_z  \vsp\\
 +\sqrt {2nB}(\omega\cos \theta -\omega '\cos \theta ') \left[ {\Lambda
_{\ell-1,n}(\beta ) \Lambda _{n-1,0}(\beta ')\varepsilon _-\varepsilon
'^*_--\Lambda_{\ell -1,n-1}(\beta ) \Lambda _{n,0}(\beta ')\varepsilon
_z\varepsilon '^*_z}\right]  \nonumber \vsp\\
 -\sqrt{2\ell B}\left\{
\begin{array}{l}
-\omega '\cos \theta '\left[ {\Lambda _{\ell ,n}(\beta )\Lambda
_{n,0}(\beta ')\varepsilon _z \varepsilon'^*_z-\Lambda _{\ell
,n-1}(\beta )\Lambda _{n-1,0}(\beta ')\varepsilon_+ \varepsilon '^*_-}
\right] \\ +\sqrt {2nB}\left[ {\Lambda _{\ell,n}(\beta )\Lambda
_{n-1,0}(\beta ')\varepsilon _z\varepsilon '^*_- +\Lambda_{\ell
,n-1}(\beta )\Lambda _{n,0}(\beta ')\varepsilon _+\varepsilon
'^*_z}\right]
\end{array}
\right\}
\end{array}
\right\} , \nonumber
\end{eqnarray}
and can be obtained from Eq.~(\ref{eq:F1def}) via the crossing symmetry
in Eq.~(\ref{eq:cross-symm}).

\newpage 

\begin{figure}
\epsscale{0.7} 
\plotone{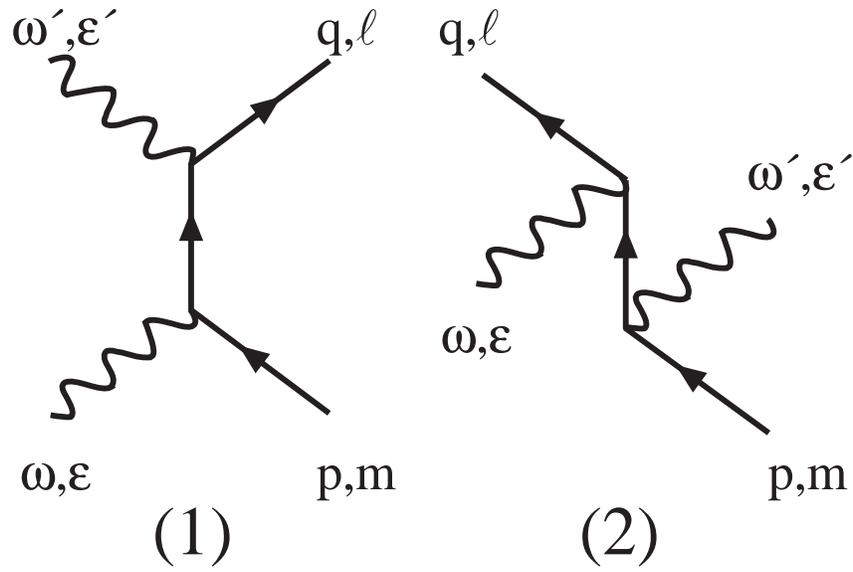} 
\caption{Feynman diagrams for Compton scattering.  Solid lines represent
electron wavefunctions scattering in a magnetic field from an initial
Landau state with quantum number, $m$, and parallel momentum, $p$, to a
final Landau state with quantum number, $\ell$, and parallel momentum,
$q$.  The wavy lines represent photon wavefunctions with an initial
energy, $\omega$, and polarization, $\varepsilon$, scattering to a final
energy, $\omega '$, and polarization, $\varepsilon '$.}
\end{figure}


\newpage
\begin{figure}
\epsscale{1.0} 
\plotone{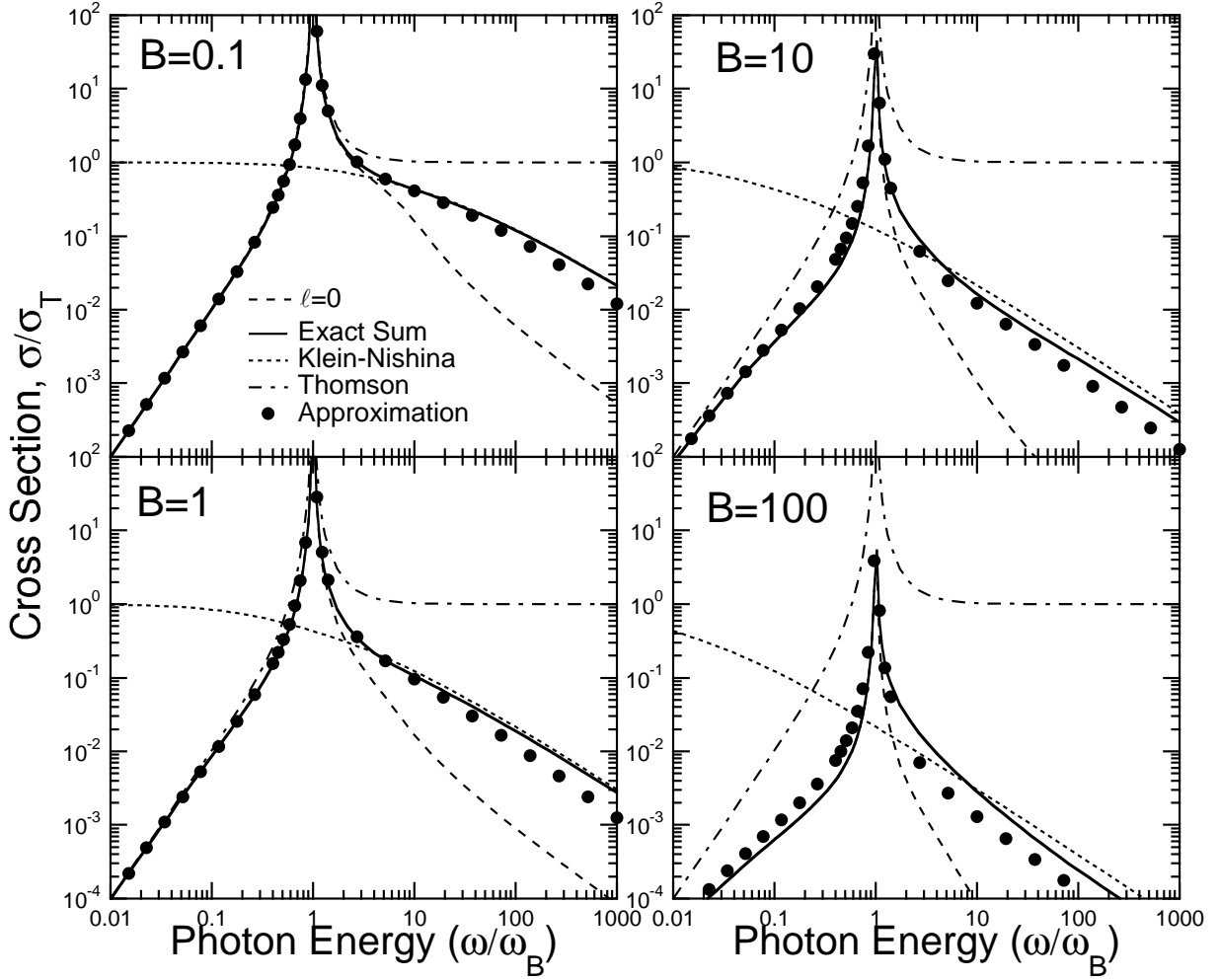} 
\caption{ Total Compton scattering cross section (in Thomson units) as a
function of the incident photon energy (in units of the cyclotron
energy) for the indicated magnetic field strengths (in units of the
critical field, $B_{\rm cr}$).  The exact QED scattering cross section,
summed over all contributing final electron Landau states is indicated
as a solid curve.  The non-relativistic magnetic Thomson cross section
is plotted as a dot-dashed curve (labeled Thomson), while the
Klein-Nishina cross section is plotted as a dotted curve.  The cross
section for only the final Landau state $\ell=0$ is plotted as a dashed
curve. }
\end{figure}

\newpage
\begin{figure}
\epsscale{1.0} 
\plotone{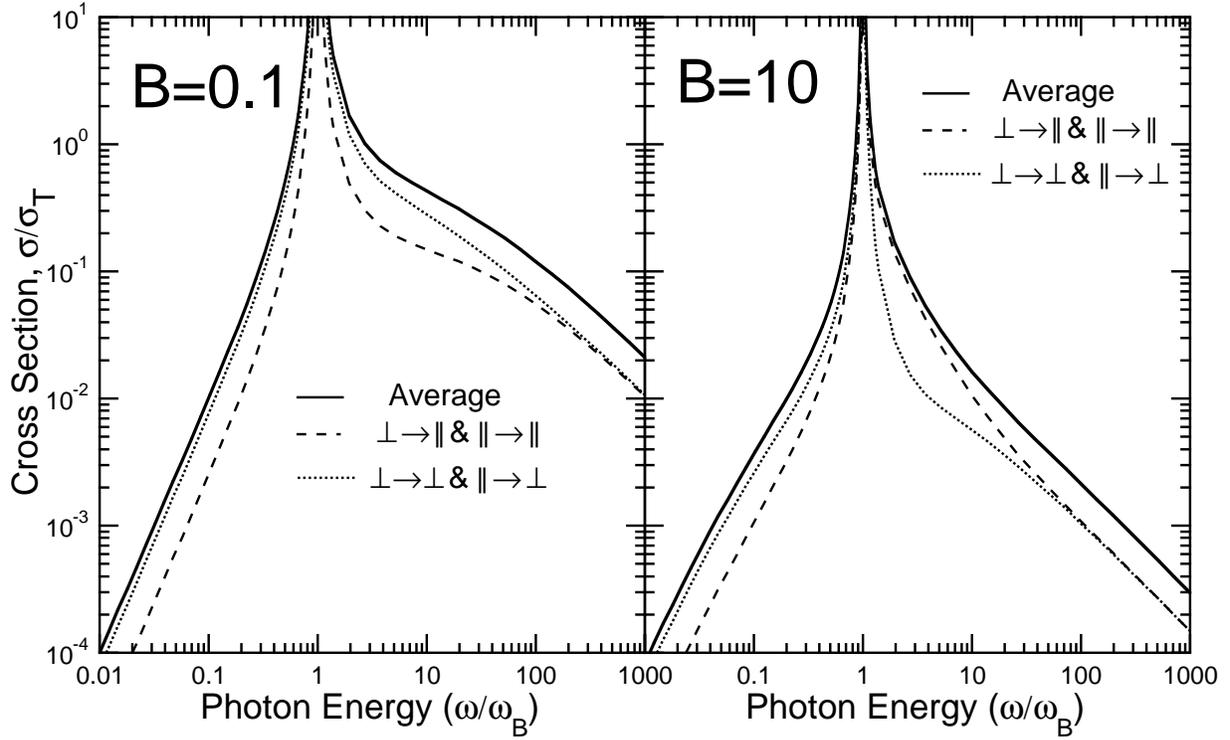} 
\caption{ Total Compton scattering cross section (in Thomson units) as a
function of the incident photon energy (in units of the cyclotron
energy) for the indicated magnetic field strengths in units of the
critical field, $B_{\rm cr}$.  The exact QED scattering cross section,
summed over all contributing final electron Landau states and averaged
over photon polarization states is indicated as a solid curve.  The QED
cross section leading to parallel and perpendicular polarizations are
plotted as dashed and dotted curves, respectively. }
\end{figure}

\newpage
\begin{figure}
\epsscale{0.75} 
\plotone{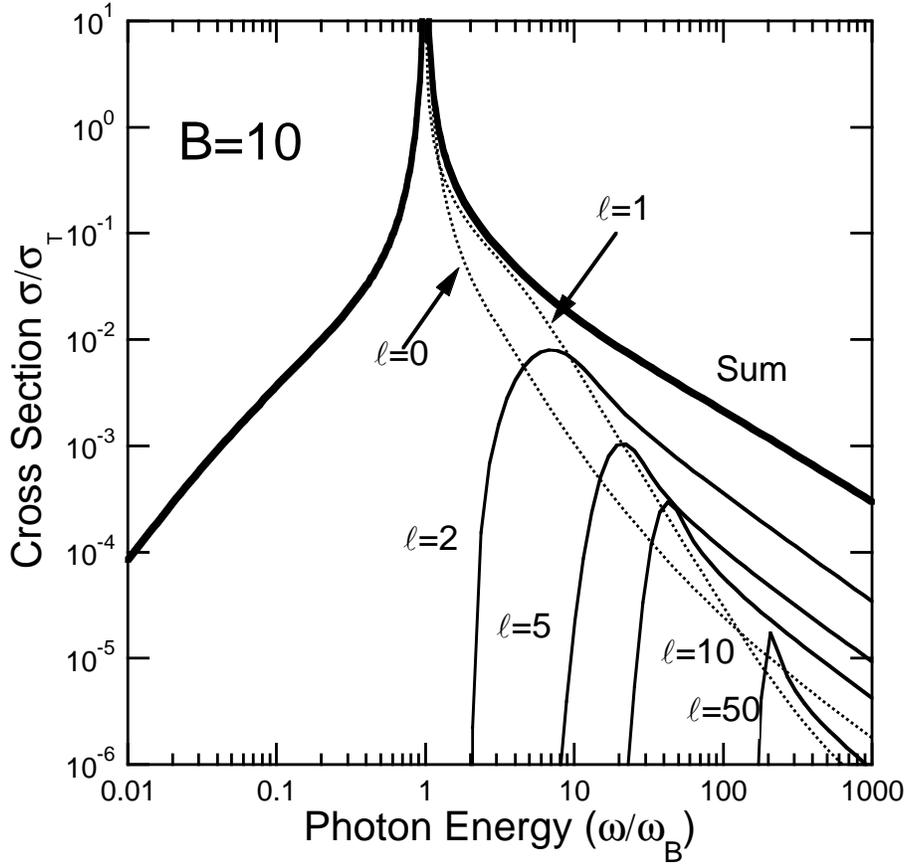} 
\caption{ Total Compton scattering cross section (in Thomson units) as a
function of the incident photon energy (in units of the cyclotron
energy) for a magnetic field strength of 10 times the critical field,
$B_{\rm cr}$.  The exact QED scattering cross section, summed over all
contributing final electron Landau states is indicated as a dark solid
curve.  The cross section for a select group of final Landau states
indicated is plotted for $\ell=0$ and $\ell=1$ as dotted curves, while
higher Landau states are plotted as light solid curves. }
\end{figure}

\newpage
\begin{figure}
\epsscale{1.0} 
\plotone{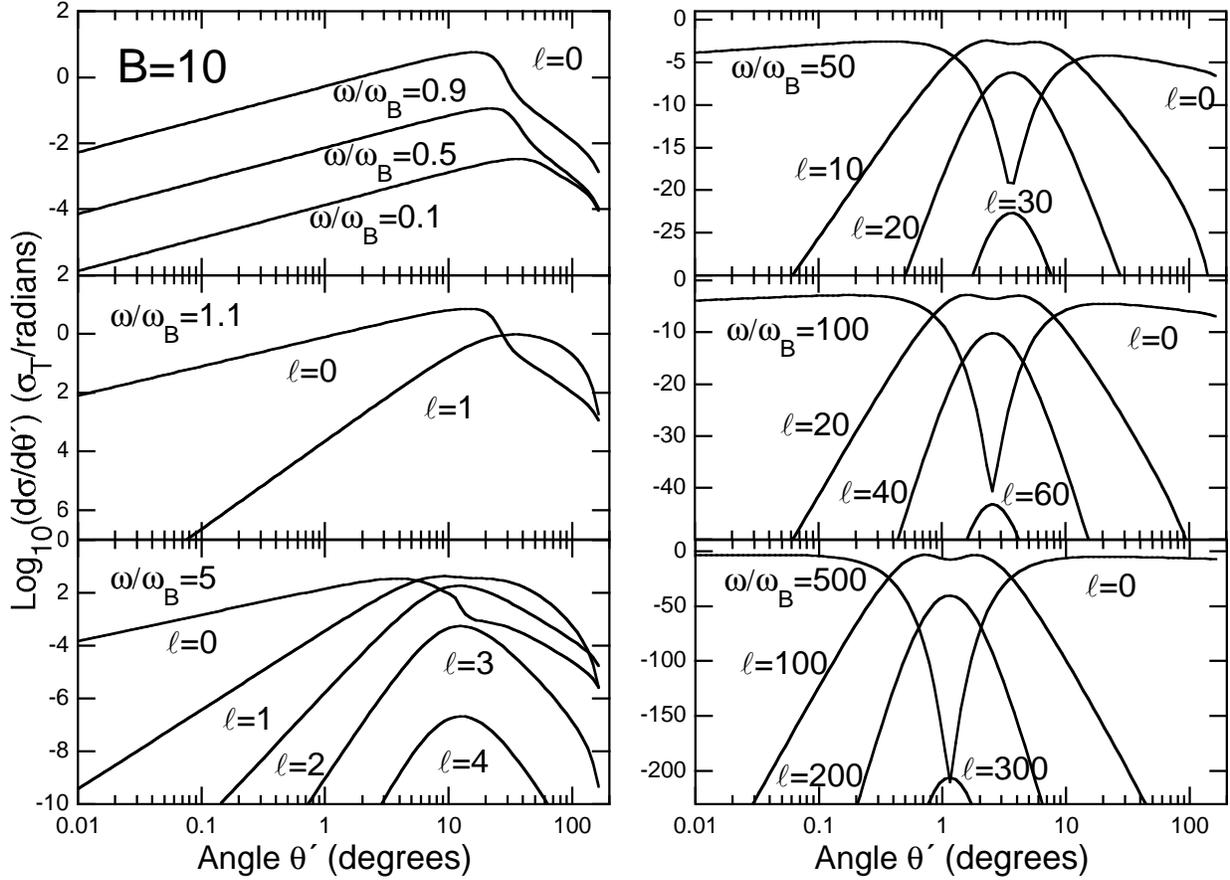} 
\caption{ The logarithm of the angular distribution, $d\sigma/d\theta'$,
for the exact QED Compton scattering as a function of the scattered
photon angle for a magnetic field strength of 10 times the critical
field, $B_{\rm cr}$.  Angular distributions are plotted for the indicated
photon incident energies and final Landau states. }
\end{figure}

\newpage
\begin{figure}
\epsscale{1.0} 
\plotone{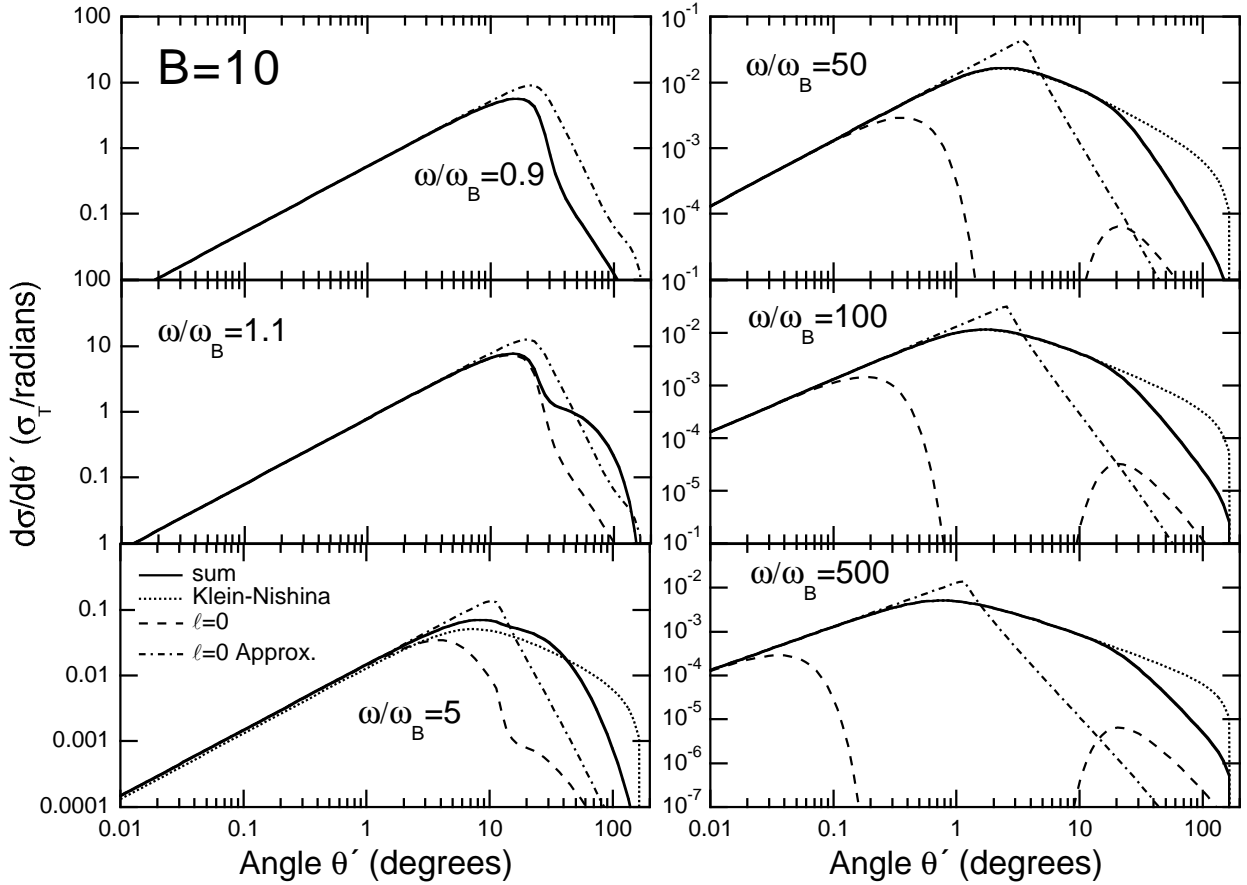} 
\caption{ The angular distribution, $d\sigma/d\theta'$, for the exact
QED Compton scattering as a function of the scattered photon angle for a
magnetic field strength of 10 times the critical field, $B_{\rm cr}$. The
exact QED angular distribution, summed over all contributing final
electron Landau states is indicated as a solid curve, while the
Klein-Nishina angular distribution is plotted a dotted curve.  The
contributions for the $\ell = 0$ final Landau state calculated using
the exact expression of equation (\ref{eq:dsig_leq0_exact}) and
calculated using the approximate expression of equation
(\ref{eq:dsig_leq0_approx}) are plotted as dashed and dot-dashed
curves, respectively.}
\end{figure}

\newpage
\begin{figure}
\epsscale{0.75} 
\plotone{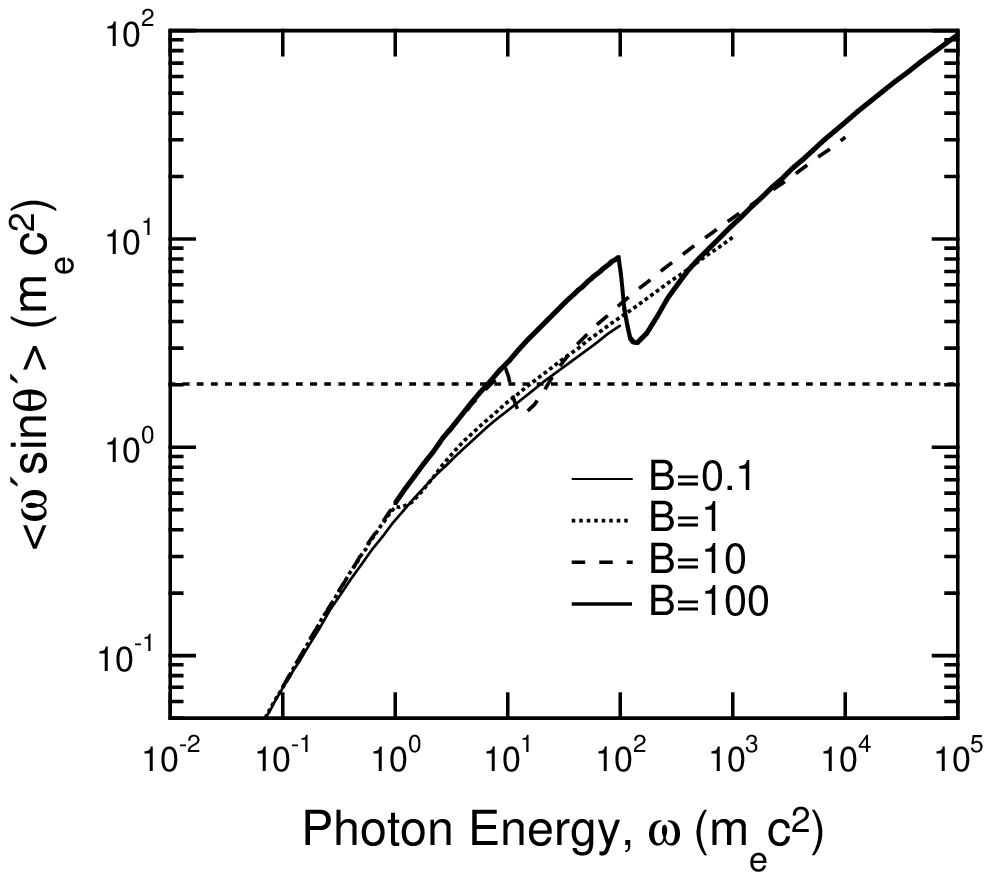} 
\caption{The quantity $\omega\sin\theta'$, in $m_ec^2$ units, averaged
over the all contributing final Landau states, $\ell$, and over the
photon scattered angle, $\theta'$, is plotted as a function of the
incident photon energy, $\omega$, in $m_ec^2$ units for the indicated
magnetic field strengths in units of the critical field $B_{\rm cr}$.  The
horizontal dashed line represents the magnetic single photon pair
creation threshold.}
\end{figure}

\begin{references} 

\reference{}
   Adler, S. L. 1971, Ann. Phys., 67, 599.
\reference{}
   Alexander, S. G.\& Meszaros, P. 1989, ApJ, 344, L1.
\reference{}
   Alexander, S. G.\& Meszaros, P. 1991, ApJ, 372, 565. 
\reference{}
   Araya, R. A. \& Harding, A. K. 1996, ApJ, 463, L33.
\reference{}
   Araya, R. A. \& Harding, A. K. 1999, ApJ, 517, 334.
\reference{}
   Baring, M. G. 1991, \aap, 249, 581.
\reference{}
   Baring, M. G. 1995, \apj, 440, L69.
\reference{}
   Baring, M. G. \& Harding, A. K. 1998, \apj, 507, L55.
\reference{}  
   Blandford, R. D. \& Scharlemann, E. T. 1976, \mnras, 174, 59.
\reference{}
   Bussard, R. W., Alexander, S. B., \& Mészáros, P. 1986, \prd, 34, 440.
\reference{}
   Camilo, F. et al. 1999, \nat, submitted.
\reference{}
   Canuto, V., Lodenquai, J. \& Ruderman, M., 1971, \prd, 3, 2303.
\reference{}
   Daugherty, J. K. \& Bussard, R. W. 1980, \apj, 238, 296.
\reference{}
   Daugherty, J. K., \& Harding, A. K. 1986, \apj, 309, 362. (DH86)
\reference{}
   Daugherty, J. K., \& Harding, A. K. 1989, \apj, 336, 861.
\reference{}
   Dermer, C. D. 1990, \apj, 360, 197.
\reference{}
   Gonthier, P. L., \& Harding, A. K. 1994, \apj, 425, 747.
\reference{}
   Graziani, C. 1993, \apj, 412, 351.
\reference{}
   Graziani, C., Harding, A. K. \& Sina, R. 1995, \prd, 51, 7097.
\reference{}
   Harding, A. K. \& Daugherty, J. K. 1991, \apj, 374, 687.
\reference{}
   Harding, A. K., Baring, M. G. \& Gonthier, P. L. 1996, Astron.
   \& Astr. Supp., 120(4), 111.
\reference{}
   Harding, A. K., Baring, M. G. \& Gonthier, P. L. 1997, \apj, 476, 246.
\reference{}
   Harding, A. K. \& Muslimov, A. G. 1998, \apj, 508, 328.
\reference{}
   Herold, H. 1979, \prd, 19, 2868.
\reference{}
   Herold, H., Ruder, H. \& Wunner, G. 1982, \aap, 115, 90.
\reference{}
   Hurley, K. et al. 1999, \nat, 397, 41.
\reference{}
   Johnson, M. H., \& Lippmann, B. A. 1949, Phys. Rev., 76, 828.
\reference{}
   Kouveliotou, C.  et al. 1998a, \nat, 393, 235.
\reference{}
   Kouveliotou, C.  et al. 1998b, \iaucirc 7001.
\reference{}
   Kouveliotou, C. et al. 1999, \apj, 510, L115.
\reference{}
   Kulkarni, S. R., \& Thompson, C.  1998, \nat, 393, 215.
\reference{}
   Latal, H. G. 1986, \apj, 309, 372.
\reference{}
   Luo, Q. 1996, \apj, 468, 338.
\reference{}
   Mereghetti, S., Caraveo, P. \& Bignami, G. F. 1992,\aap, 263, 172.
\reference{}
   Mereghetti, S. \& Stella, L. 1995, \apj, 442, L17.
\reference{}
   Meszaros, P. 1992, High-energy radiation from magnetized neutron stars,
   (Chicago: Univ. of Chicago Press).
\reference{}
   Miller, M. C. 1995, \apj, 448, L29.
\reference{}
   Paczynski, B. 1992, Acta Astro., 42, 145.
\reference{}
   Rybicki, G.  B. \& Lightman, A. P., 1979, {\it Radiative Processes
   in Astrophysics}, (New York, Wiley-Interscience), p~195.
\reference{}
   Shabad, A. E. 1975, Ann. Phys., 90, 166.
\reference{}
   Shapiro, S. L. \& Teukolsky, S. A., 1983, {\it Black Holes, White
   Dwarfs, and Neutron Stars The Physics of Compact Objects} (John Wiley \&
   Sons New York), 278.
\reference{}
   Shitov, Yu. P. 1999, IAU Circ., 7110, 2.
\reference{}
   Shitov, Yu. P., Pugachev, V.~D. \& Kutuzov, S.~M. 2000, in Pulsar
   Astronomy: 2000 and Beyond (IAU Coll. 177), eds. M. Kramer, N. Wex
   \& R. Wielebinski (ASP Conf. Ser., San Francisco)
   [{\tt astro-ph/0003042}]
\reference{}
   Sina, R. 1996, Ph.D. Thesis University of Maryland.
\reference{}
   Steinle, H., Pietsch, W., Gottwald, M. \& Graser, U. 1987,
   \aaps, 131, 687.
\reference{}
   Sturner, S. J. 1995, \apj, 446, 292.
\reference{}
   Taylor, J. H., Manchester R. N., and Lyne, A. G. 1993, \apjs,
   88, 529, (see also {\tt http://pulsar. princeton.edu}).
\reference{}
   Thompson, C. \& Duncan, R. C. 1995, \mnras 275, 255.
\reference{}
   Thompson, C. \& Duncan, R. C. 1996, \apj, 473, 332.
\reference{}
   Usov, V. V. \& Melrose, D. B. 1995, Aust. J. Phys., 48, 571.
\reference{}
   Vasisht, G. \& Gotthelf, E. V. 1997, \apj, 486, L129.
\reference{}
   Xia, X. Y., Qiao, G. J., Wu, X. J., \& Hou, Y. Q. 1985, \aap, 152, 93.
\reference{}
   Zhang, B., Qiao, G. J., Lin, W. P. \& Han, J. L. 1997, \apj, 478, 313.

\end{references}
\end{document}